\documentclass[12pt,onecolumn,draftcls]{IEEEtran}

\pdfoutput=1

\usepackage{color}
\usepackage{float}
\usepackage{url}
\usepackage{amsmath,epsfig}
\usepackage{amssymb}
\usepackage{times}
\usepackage{graphicx}
\usepackage{theorem}
\usepackage{cite}
\usepackage{euscript}
\usepackage{pifont}
\usepackage{fancyhdr}
\usepackage{subfigure}
\usepackage{setspace}

\newcommand{\R}{\mathbb{R}}
\newcommand{\N}{\mathcal{N}}
\newcommand{\Es}{E_{\text{s}}}
\newcommand{\Eb}{E_{\text{b}}}
\newcommand{\Pout}{P_{\rm out}}
\renewcommand{\Pr}{{\mathbb P}}
\newcommand{\Ic}{\mathcal{I}}
\DeclareMathOperator{\tthh}{\text{th}}
\DeclareMathOperator{\sign}{\text{sign}}
\newcommand{\alphav}{\hbox{\boldmath$\alpha$}}

\newtheorem{definition}{Definition}
\newtheorem{proposition}{Proposition}
\newtheorem{corollary}{Corollary}
\newfloat{floateq}{!b}{loe}[section]

\title{Low-Density Parity-Check Codes for Nonergodic Block-Fading Channels}

\author{Joseph J.\ Boutros, Albert Guill{\'e}n i F{\`a}bregas,
Ezio Biglieri, and Gilles Z\'emor%
\thanks{Joseph J.\ Boutros (Email: {\tt boutros@ieee.org}) is with
Texas A{\&}M University at Qatar, Doha, Qatar. Albert Guill{\'e}n
i F{\`a}bregas (Email: {\tt guillen@ieee.org}) is with the Department of Engineering.
University of Cambridge, Cambridge, UK. Ezio Biglieri (Email: {\tt
e.biglieri@ieee.org}) is with DITIC, Universitat Pompeu Fabra,
Barcelona, Spain. Gilles Z\'emor (Email: {\tt
zemor@math.u-bordeaux1.fr}) is with the Institut de Math\'ematiques de Bordeaux, Universit\'e de Bordeaux 1, Bordeaux, France.}
\thanks{This work has been presented in part at the 2007 Information Theory and Applications Workshop, San Diego, CA, USA, January-February 2007.}
\thanks{The work of Ezio Biglieri was supported by the Spanish Ministery
of Education and Science under Project TEC2006-01428/TCM.} }
\begin{document}
\maketitle
\begin{abstract}
We solve  the problem  of designing powerful  low-density
parity-check (LDPC) codes with iterative decoding for the
block-fading channel.  We first study the case of maximum-likelihood decoding, and show that the design criterion is rather straightforward. Unfortunately, optimal constructions for maximum-likelihood decoding do not perform well under iterative decoding. To overcome this limitation, we then introduce a new family of full-diversity
LDPC codes that  exhibit near-outage-limit  performance under iterative decoding for all block-lengths.  This
family  competes  with  multiplexed parallel turbo codes suitable
for nonergodic channels and recently reported in the literature.
\end{abstract}

\newpage
\section{Introduction}
The  block-fading (BF) channel model was first  introduced in
\cite{Ozarow1994}, and further elaborated upon in
\cite{Biglieri1998} (see also~\cite[p.\ 98 ff.]{Biglieri2005}).
This is a realistic and convenient model for a number of channels
affected by slowly varying fading, and, as observed for example
in~\cite{Guillen2006-a}, is especially relevant in wireless
communications involving slow time--frequency hopping (e.g.,
cellular networks and wireless Ethernet) or multicarrier
modulation using orthogonal frequency division multiplexing
(OFDM). The design of error-control codes for BF channels offers a
challenging problem, which differs greatly from its counterparts
referred to additive white Gaussian noise (AWGN) or
independent-fading channels (see~\cite{Guillen2006-a} for a
summary of recent results). The main reason for this unlikeness
stems from the fact that in BF channels the random channel gains
remain constant during a block of symbols (see below for
additional details and definitions), and take independent values
from block to block. As a result, while the word-error probability
in independent-fading channels depends on the Hamming distances
between code words, in BF channels it depends on a new parameter,
the {\em blockwise Hamming distance}. Since codes exhibiting a
large minimum Hamming distance may not have a large blockwise
Hamming distance, codes that are good when used on the
independent-fading channel may not be as good for a BF channel. In
addition, over independently faded channels permutations of the
symbols cause no variation of the code performance, but this
property does not hold on the BF channel. Thus, if an
off-the-shelf code, designed for the independent-fading channel,
is used for transmission over the BF channel, it is important to
carefully select the best permutation of its symbols. Finally, one must
consider that the BF channel is nonergodic. As a consequence, to
determine the information-theoretical rate limit which cannot be
surpassed by the word error probability of any coding scheme, one
cannot use channel capacity, but rather {\em outage
probability}~\cite{Ozarow1994,Biglieri1998,Biglieri2005}.
Classical random-like codes, designed to approach ergodic
capacity, cannot generally approach the ideal performance limits
of BF channels, and hence code designs suited to the nonergodic nature of the channel are called
for. This paper is devoted to this design problem.

Two  main  parameters that determine  the  error rate of coded BF
channels for high signal-to-noise (SNR) ratios are the {\em
diversity order} and the {\em coding gain}. The former determines
the slope of the error-rate curve as a function of the SNR on a
log-log scale\footnote{The diversity order
 is exactly the asymptotic slope for
Rayleigh fading, while for other fading distributions it is only
proportional to the slope. See~\cite{wangia,Nguyen2007} for details. In this
paper we shall restrict our attention to Rayleigh fading.}.
Since the error probability of any coding scheme is lower-bounded
by the outage probability, the diversity order is upper-bounded by
the {\em intrinsic diversity} of the channel, which reflects the
slope of the outage limit. When maximum diversity is achieved by a
code, the coding gain yields a measure of SNR proximity to the
outage limit. The maximum achievable diversity order with discrete
input constellations is given by the Singleton
bound~\cite{Knopp2000,Malkamaki1999,Guillen2006-a}, and codes
achieving the Singleton bound are termed blockwise
maximum-distance separable (MDS). Blockwise MDS codes are
outage-achieving  over the (noiseless) block-erasure
channel~\cite{Guillen2006-c}, but may not achieve the
outage-probability limit on noisy BF channels. 
As a matter of fact, as shown in \cite{Guillen2006-a}, blockwise MDS codes are necessary, but not sufficient to approach the outage probability of the channel. 

Recent code designs for BF channels include near-outage schemes
based on a suitable permutation of parallel turbo
codes~\cite{Allerton2004,Allerton2005,ITA2006}.  Multiplexers for
convolutional, turbo and repeat-accumulate codes~\cite{Allerton2004, Knopp2000,Guillen2006-a}  appeared one
decade after the  analysis of random  and  periodic  interleaving
of  convolutional  codes on the block-erasure
channel~\cite{Lapidoth1994}.   Random ensembles of low-density
parity-check codes (LDPC) designed for ergodic  AWGN
channels~\cite{Richardson2001-b,Hou2001}, in spite of the
excellent  decoding threshold  of their irregular structures, do
not have full-diversity,  and hence exhibit a poor performance over a
BF channel. Decoding thresholds of LDPC code ensembles over ergodic BF
channels have been studied \cite{Jin2004}. Unfortunately, these codes
are not designed to be blockwise MDS, and therefore fail to 
achieve the outage limit in the nonergodic setup.

In this work, we introduce a new  family of blockwise MDS LDPC codes, the {\em root LDPC codes}, based on a
special type of checknode
 that we call {\em rootchecks}. Under
iterative message-passing decoding, they  achieve the
outage-probability limit on block-erasure channels, and they
perform close to that limit on Rayleigh BF channels. This paper is
organized as follows. Section~\ref{sectionII} introduces the
channel model and the relevant notations.  LDPC codes with full
diversity under Maximum Likelihood (ML) decoding are discussed in
section~\ref{sectionIII}. Our new family of LDPC codes suited for
iterative decoding is further described. Section \ref{sectionV} analyzes their
density evolution in the presence of block fading. Conclusions are
finally drawn in Section \ref{section:conclusions}. Complementary support material is shown in the Appendix.

\section{Channel model and notation}\label{sectionII}
\begin{figure}[t!]
\begin{center}
\includegraphics[width=0.9\columnwidth]{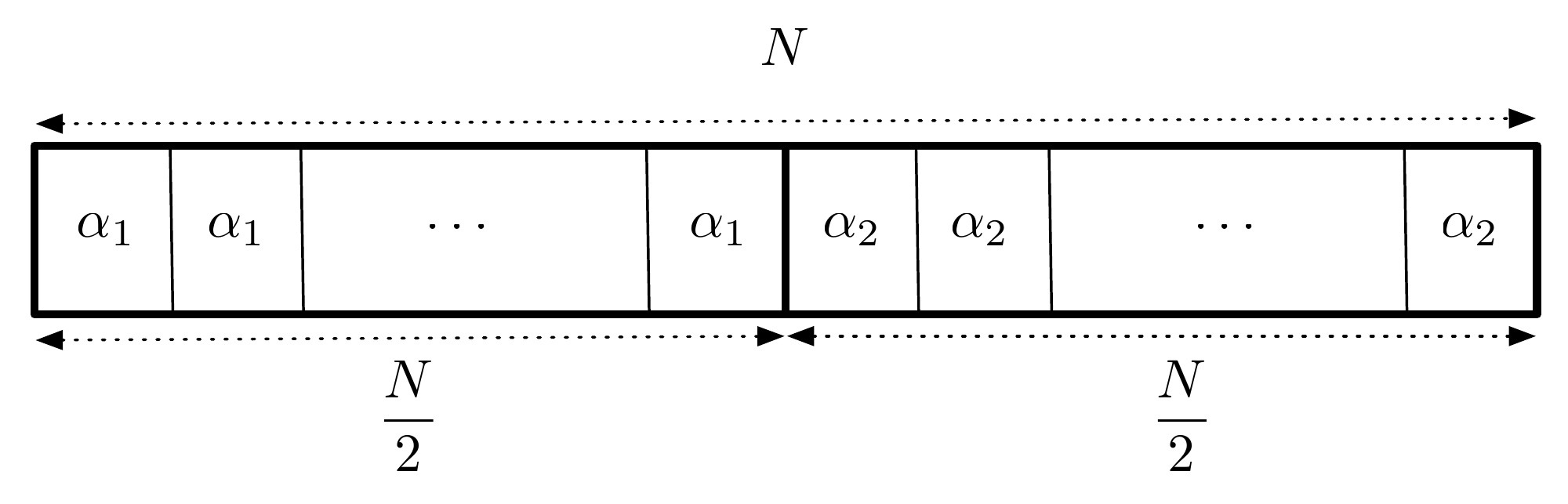}
\end{center}
\vspace{-5mm}
\caption{\sl Codeword representation for a BF channel with
$n_c=2$. The fading gains $\alpha_1$, $\alpha_2$ are independent
between themselves and among codewords.}
\label{fig_2channel_states}
\end{figure}

We consider codewords of $N$ binary digits transmitted on a BF
channel, where $n_c$ independent fading gains (whose values form
the {\em channel state}) affect each codeword. The length $N$ is a
multiple of $n_c$, with $\ell\triangleq N/n_c$ denoting the number
of bits per fading block. The received signal when symbol $x_i$ is
transmitted is given by
\begin{equation}
y_i = \alpha_j x_i + z_i
\end{equation}
where $y_i \in \R$, $i=1 \ldots N$, and $j=1+[(i-1)/\ell]$, with
$[r]$ denoting the integer part of a real number $r$.  The nonnegative
real number $\alpha_j$  is the  fading gain  at block $j$, $j=1
\ldots n_c$.  The symbols $x_i$ are chosen from a BPSK alphabet,
$x_i=\pm \sqrt{E_s}$,   where  $E_s$   is  the  average energy per
symbol. The   noise    samples   are    i.i.d.\ with $z_i \sim
\N(0,\sigma^2)$, $\sigma^2=N_0/2$. We assume perfect channel state
information (CSI)  at the receiver,  and channel gains which are
i.i.d.\ Rayleigh-distributed from block to block and from codeword
to codeword. Thus,  when the information rate is $R$ bits per
channel use, the average SNR per symbol is given by $\gamma =
\Es/N_0$, and the  average SNR per  bit is  $ \Eb/N_0  =
\gamma/R$. Fig. \ref{fig_2channel_states} illustrates the channel
model for $n_c=2$ and $\ell=N/2$.

\begin{figure}[t!]
\begin{center}
\includegraphics[width=0.8\columnwidth]{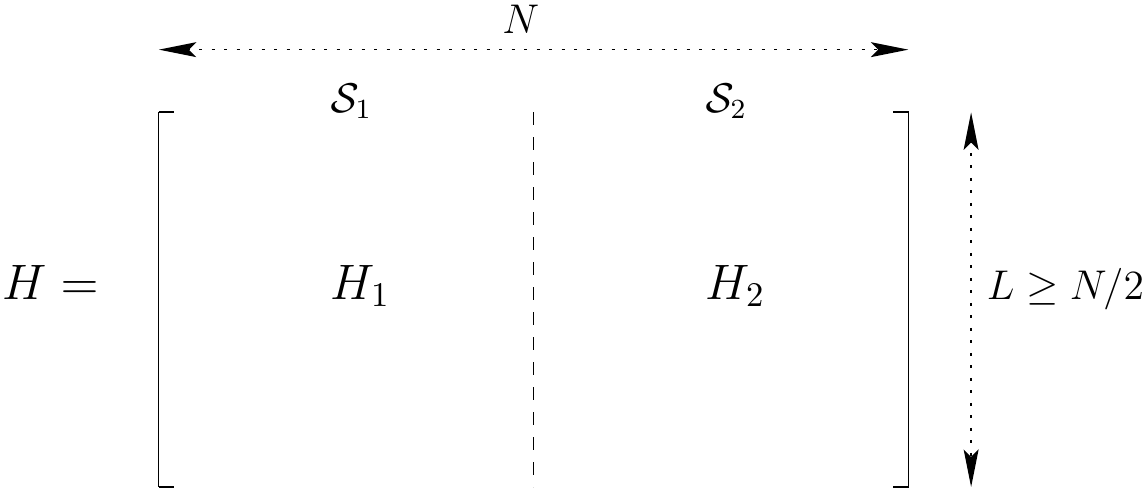}
\end{center}
\caption{\sl Parity-check matrix notations for a block-fading
channel with $n_c=2$. The $L-N/2$ extra rows are added in order to
enhance the coding gain of a full-diversity code.}
\label{fig_matrix_splitted}
\end{figure}

In this work, we focus on linear binary codes ${\EuScript
C}(N,K)_2$ with block length $N$, dimension $K$, and rate $R=K/N
\le 1/n_c \le 1/2 $. The code $\EuScript C$ is defined by an $L
\times N$ parity-check matrix $H$
(Fig.~\ref{fig_matrix_splitted}), or, equivalently, by the corresponding Tanner
graph~\cite{Biglieri2005}. This has $L$ single-parity checknodes.
It is assumed that $H$ has full rank $L$, so that $R=1-L/N$.

For a given nonzero codeword $c \in \EuScript C$, we define the
blockwise Hamming weight vector $(\omega_1,\dotsc,\omega_{n_c})$,
where $\omega_j$ is the Hamming weight of the coded bits affected
by fading $\alpha_j$. Following \cite{Knopp2000,Guillen2006-a} we define the {\em block diversity} of $\EuScript C$ as
\[
d = \min_{ c \in {\EuScript C}-\{0\}} |\{ \omega_j\neq 0 \}|.
\]
In words, the block diversity is the minimum number of blocks that
have non-zero Hamming weight, or the blockwise Hamming
distance. Qualitatively, this implies that an ML decoder of $\EuScript
C$ will be able to decode correctly in presence of $d-1$ deep fades, which one can think
  of as block erasures.

We also define the minimum blockwise Hamming
weight as
\[
\omega^\star = \min_{ c \in {\EuScript C}-\{0\}}
(\omega_1,\dotsc,\omega_{n_c}).
\]

\begin{definition}
An  error-correcting  code  is  said  to have {\em full diversity}
if $\omega^\star > 0$.
\end{definition}

Having $\omega^\star > 0$ implies that
$d=n_c$, having nonzero weight in all blocks. 

Now, observe that the  blockwise Singleton
bound~\cite{Malkamaki1999,Knopp2000,Guillen2006-a,Biglieri2005}
\[
d \leq 1+\lfloor n_c(1-R)\rfloor
\]
determines $R=1/n_c$ as the highest achievable rate for a
full-diversity code. Furthermore, the word  error  probability of
a code with diversity $n_c$ decreases as $1/\gamma^{n_c}$ at high
SNR~\cite{Proakis2000,Biglieri2005,wangia}.

The block-fading  channel is not  information
stable~\cite{Verdu1994}, and  therefore its  Shannon  capacity  is
zero since  there  is  a non-vanishing
 probability that the decoder
makes a {\em word error}. In the  limit  of  large  block  length,
this  probability  is  the  {\em information        outage
probability}, defined as~\cite{Ozarow1994,Biglieri1998}
\begin{equation}
\Pout(\gamma,R) \triangleq \Pr \{\Ic(\gamma,\alphav) < R\}
\end{equation}
where  $\Ic(\gamma,\alphav)$ is  the {\em  instantaneous
input--output mutual information} between  the input and output of
the channel, defined as
\begin{equation}
\Ic(\gamma,\alphav)  \triangleq  \frac{1}{n_c}\sum_{i=1}^{n_c}
I_{\rm AWGN}(\gamma\alpha_i^2), \label{eq:mi}
\end{equation}
with $I_{\rm AWGN}(s)$ the input--output  mutual information of an
AWGN channel  with SNR  per  symbol equal to $s$. The BF channel
is also commonly referred  to as  {\em nonergodic} since, for
finite values of $n_c$, $\Ic(\gamma,\alphav)$ is a non-constant
random variable.

The  information outage  probability $\Pout(\gamma,R)$  is a
fundamental lower bound on
the  word   error   rate  for  sufficiently large
word length.  Therefore, any code  approaching $\Pout(\gamma,R)$
should have  a word-error probability that, as $N$ increases,
becomes   {\em   independent}    of the    code
length~\cite{Allerton2005,Guillen2006-a}.

Unless stated otherwise, we shall focus our study on a coding rate
$R=\frac{1}{2}$ (or just slightly smaller than $\frac{1}{2}$) and a nonergodic
Rayleigh fading channel with $n_c=2$ blocks per codeword, as
depicted in Figs.~\ref{fig_2channel_states} and
\ref{fig_matrix_splitted}. However, most of our results can be
easily generalized to $R=\frac{1}{n_c}$.

\section{Full-diversity LDPC codes under ML
decoding}\label{sectionIII}
In this section, we study LDPC codes in the presence of BF under
ML decoding. As we shall see, the design of full-diversity LDPC
codes under ML decoding is rather straightforward. We recognize
that ML decoding is unfeasible in practice; however, it yields
valuable insight into code structures suitable for nonergodic
channels. The main result of this section is somewhat negative:
under iterative decoding, ML-designed full-diversity codes fail to
guarantee diversity, due to badly located pseudo-codewords.

Following the notations defined in the previous section, the $L
\times N$ parity-check matrix $H$ is written in the form
$H=[H_1~|~H_2]$, where the left and right parts $H_1$, $H_2$ are
$L \times N/2$. The vector space generated by the $N/2$ left
columns is denoted $\mathcal{S}_1$. Similarly, $\mathcal{S}_2$ is
the vector space generated by the $N/2$ right columns.

\begin{proposition}
A binary code $\EuScript C$ with rate $R \le \frac{1}{2}$, i.e. $L \ge
N/2$, has full diversity if and only if $H_1$ and $H_2$ are both
full-rank.
\end{proposition}
\begin{proof}
If $\dim{\mathcal{S}_1}=N/2$, then a nonzero codeword cannot have
its support on $H_1$, because all columns in $H_1$ are
independent. Hence, $\omega_2 > 0$ 
 for all nonzero codewords.
Similarly, $\omega_1 > 0$
 when $\dim{\mathcal{S}_2}=N/2$. Finally,
$\omega_1 > 0$ and $\omega_2 > 0$ for all nonzero codewords, which
yields $\omega^{\star} > 0$.
\end{proof}
\vspace{12pt}

The full-rank property of the above proposition was first observed
in~\cite{Hirst2003}. Its extension to coding rate $1/3$ with
$H=[H_1~|~H_2~|~H_3]$ can be obtained by imposing that the
matrices $[H_1~|~H_2]$, $[H_1~|~H_3]$, and $[H_2~|~H_3]$  all have
full rank. Generalization to any rate $R=\frac{1}{n_c}$ is
straightforward\footnote{Two interesting combinatorial problems
arising from this Proposition are the following: (1)~What is the
probability of a random binary matrix to be full-rank? And,
(2)~What is the probability of a random binary {\em sparse} matrix
to be full-rank?}.


\begin{proposition}
\label{prop_wmin=1} Consider a binary code $\EuScript C$ with rate
$R=1/2$, and hence with $L=K=N/2$. If $\EuScript C$ has full
diversity, then $\omega^{\star}=1$.
\end{proposition}
\begin{proof}
If $\EuScript C$ has full diversity, then
$\dim{\mathcal{S}_1}=\dim{\mathcal{S}_2}=N/2$. Any column from
$H_1$ can then be written as a linear combination of columns from
$H_2$. This is also valid for any column belonging to $H_2$.
Hence, nonzero codewords with $\omega_i=1$ exist for both $i=1$
and $i=2$ if the coding rate is exactly equal to $1/2$.
\end{proof}
\vspace{12pt}

The minimum blockwise Hamming weight must be increased in order to
improve the coding gain of $\EuScript C$.
Proposition~\ref{prop_wmin=1} 
states that to achieve this, one must decrease 
the coding rate. 
The next proposition shows that adding just one extra row
is enough to move from $\omega^{\star}=1$ to $\omega^{\star}=2$
under ML decoding.



\begin{proposition}
There exists a binary code $\EuScript C$ of rate $R=1/2-1/N$ 
that has full diversity with $\omega^\star =2$.
\end{proposition}

\begin{figure}[htb]
\begin{center}
\includegraphics[width=0.8\columnwidth]{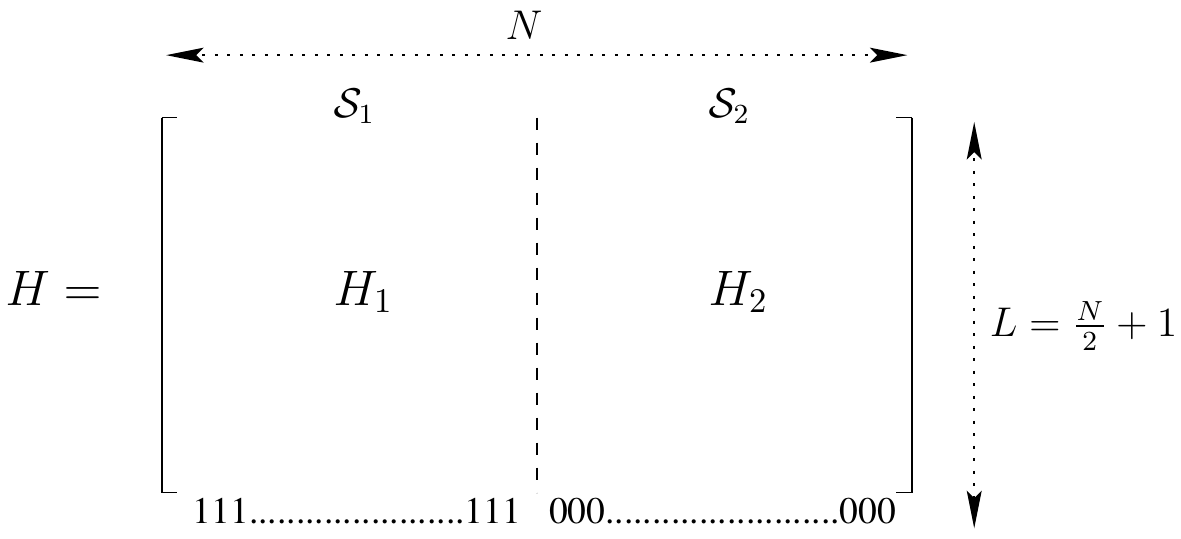}
\caption{\sl ML-designed full-diversity LDPC code with $\omega^{\star}=2$.} 
\label{fig_matrix_wmin=2_gilles}
\end{center}
\end{figure}

\begin{proof}
The proof is based on the special parity-check matrix structure
shown in Fig.~\ref{fig_matrix_wmin=2_gilles} where $H_2$ is a
full-rank matrix whose columns have odd Hamming weight 
(the identity matrix, for example). 
Let now $H_1$ be such that its first column is the all zero vector, 
and the remaining $N/2-1$ columns are all even-weight and 
full-rank.

Next, we show that the $\omega^\star$ corresponding to this
construction is $2$.
Clearly the first (leftmost) $N/2$ columns of $H$ and the last
(rightmost) $N/2$ columns of $H$ have full rank, so that
 we have $\omega^\star \geq 1$. 

None of the first $N/2$ columns of $H$ can be a linear combination of
the last $N/2$ columns of $H$, due to the $1$ in the last position of each of the first columns. 
None of the last $N/2$ columns of $H$ can be a linear combination of
the first $N/2$ columns of $H$, because columns of $H_2$ have odd
weight and any linear combination of columns of $H_1$ has even weight.

These last statements imply that $\omega^\star \geq 2$.
\end{proof}

 The rate reduction necessary to achieve
$\omega^{\star}=2$ is negligible for large code length $N$. If we
now require $\omega^{\star}=3$, the following result holds:

\begin{proposition}
\label{prop_wmin=3} Consider a binary code $\EuScript C$ with rate $R \le
1/2$. The code has $\omega^{\star}=3$ only if 
$R \le 1/2-(1/N)\log_2(1+N/2)$.
\end{proposition}

\begin{proof}
Denote by $H_2^{\rm col}$ the set of columns of $H_2$. Consider the $1+N/2$ sets consisting of $H_2^{\rm col}$ together with
its translates $h_1+H_2^{\rm col}$ for all columns $h_1$ of $H_1$.
No two of these sets can intersect, otherwise either a column of
$H_1$, or a sum of two columns of $H_1$, equals a sum of columns of
$H_2$, which would imply the existence of a codeword of weight at
most $2$ on the first $N/2$ positions. Therefore we must have
$2^L\geq (1+N/2)2^{N/2}$.
\end{proof}

\vspace{12pt}

\begin{figure}[htb]
\begin{center}
\includegraphics[width=0.85\columnwidth]{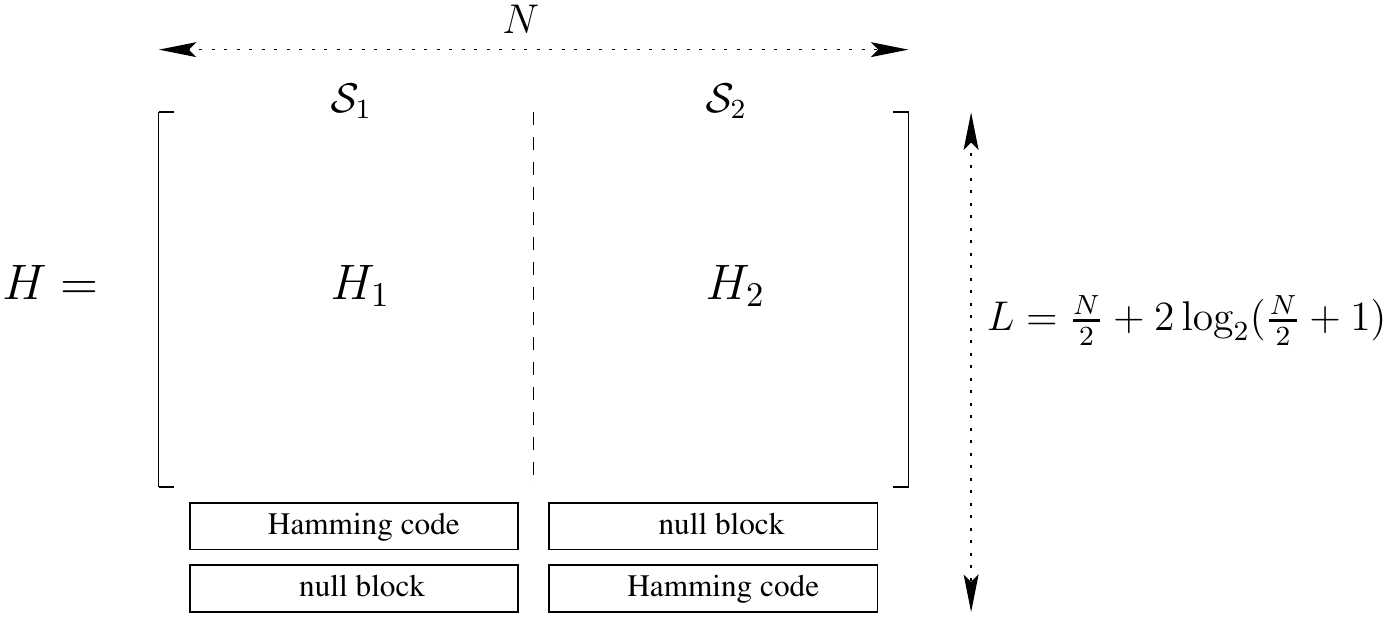}
\end{center}
\caption{\sl ML-designed full-diversity LDPC code with
$\omega^{\star} \ge 3$.} \label{fig_matrix_wmin=3}
\end{figure}

\begin{proposition}
There exists a full-diversity binary code with $\omega^{\star} \ge
3$ and $R=1/2-(1/N)2\log_2(N/2+1)$.
\end{proposition}
\begin{proof}
The  code has the parity-check matrix of
Fig.~\ref{fig_matrix_wmin=3}. The presence of a Hamming code whose
minimum distance is $3$ rules out a blockwise Hamming weight equal
to $2$.
\end{proof}
\vspace{12pt}

\begin{figure}[htb]
\begin{center}
\includegraphics[width=0.99\columnwidth]{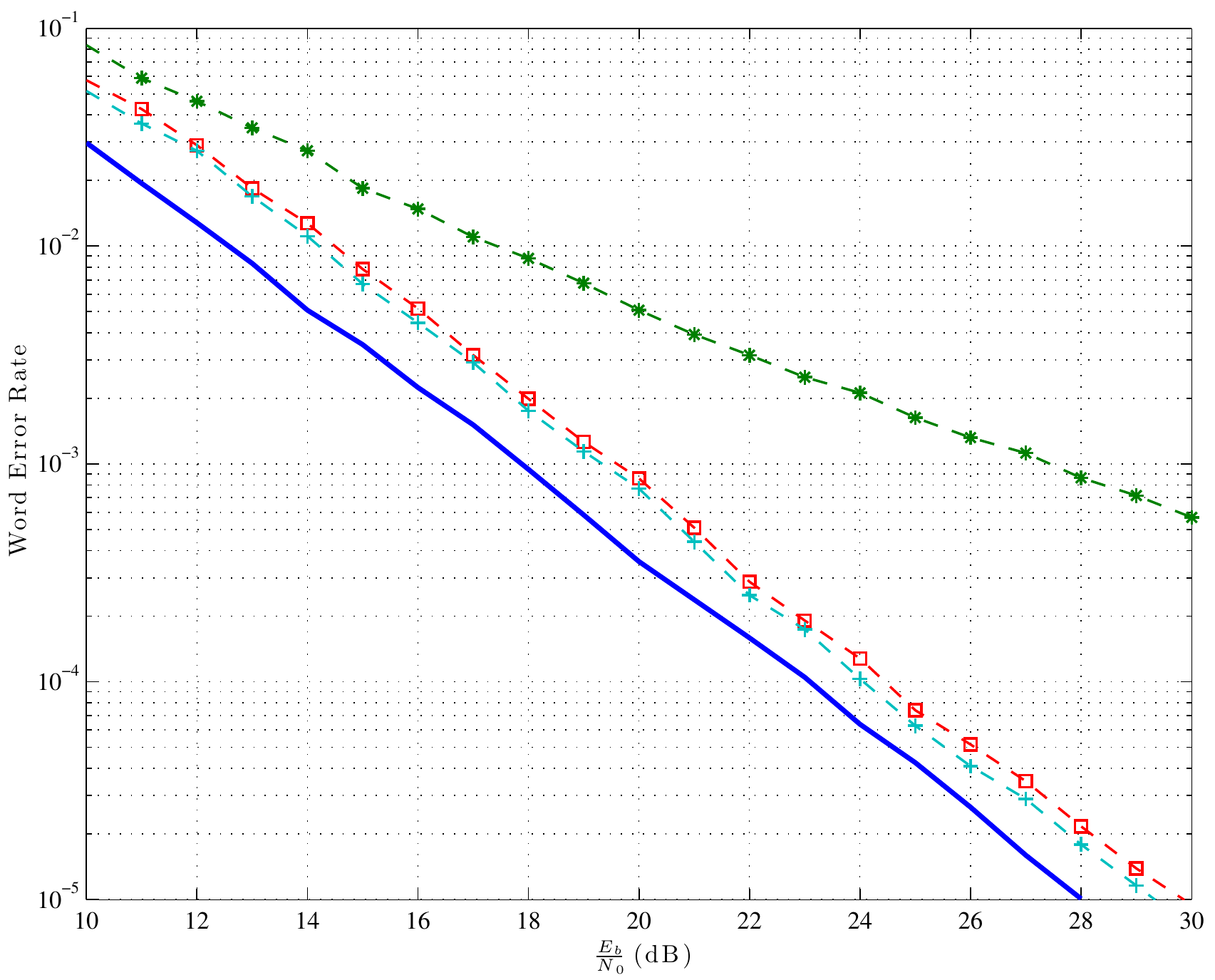}
\end{center}
\caption{\sl Rate $1/2$ ML-designed LDPC codes with iterative
decoding on a Rayleigh block-fading channel with $n_c=2$. The thick solid line corresponds to the outage probability with BPSK inputs, the dotted lines with $*$ markers corresponds to the ML-designed code with iterative decoding,  the dotted lines with $\square$ markers corresponds to the ML-designed code with $\omega^\star=1$ using a genie ML decoder and the dotted lines with $+$ markers corresponds to the ML-designed code with $\omega^\star=3$ using the genie ML decoder. The genie ML 
curves show the performance of a decoder that knows whether errors
occur in positions corresponding to $H_1$ or $H_2$.}
\label{fig_perf_ldpc_ml}
\end{figure}

We are now in a position to examine the  word-error rate of
full-diversity LDPC codes designed for ML decoding, and compare it
to the outage capacity limit. The results  are  illustrated  in
Fig.~\ref{fig_perf_ldpc_ml},  for $n_c=2$ and the  $(3,6)$
ensemble using the  constructions outlined above. With iterative
decoding, an ML-designed LDPC code has diversity one.   This
effect is caused by the pseudo-codewords~\cite{Koetter2003}  whose
support is restricted  to $H_1$  or  $H_2$, and hence have a
minimum blockwise pseudo-weight equal to zero when  iterative
belief propagation decoding is applied.  Even a random LDPC code
(not shown in the figure) performs as poorly as an ML-designed
code with $\omega^{\star}=1$. On the other hand, full diversity is
guaranteed when a ``genie-aided'' ML decoder is used which knows whether errors occur in positions corresponding to
$H_1$ or to $H_2$.

%
\section{Full-diversity LDPC codes for iterative belief propagation decoding}
\label{sectionIV}
The results presented at the end of Section~\ref{sectionIII} show
that, if iterative decoding is used, the design criteria derived
under the assumption of ML decoding are irrelevant. In this
section, we proceed to design LDPC codes with the stipulation of
iterative decoding. Our design is based on a graphical
representation~\cite{Biglieri2005,Richardson2007}, which is then
translated into a matrix description. We then analyze the construction by means of log-ratio
probability-density evolution.
\subsection{A limiting case: block-erasure channels}
\begin{figure}[b!]
\begin{center}
\includegraphics[width=0.65\columnwidth]{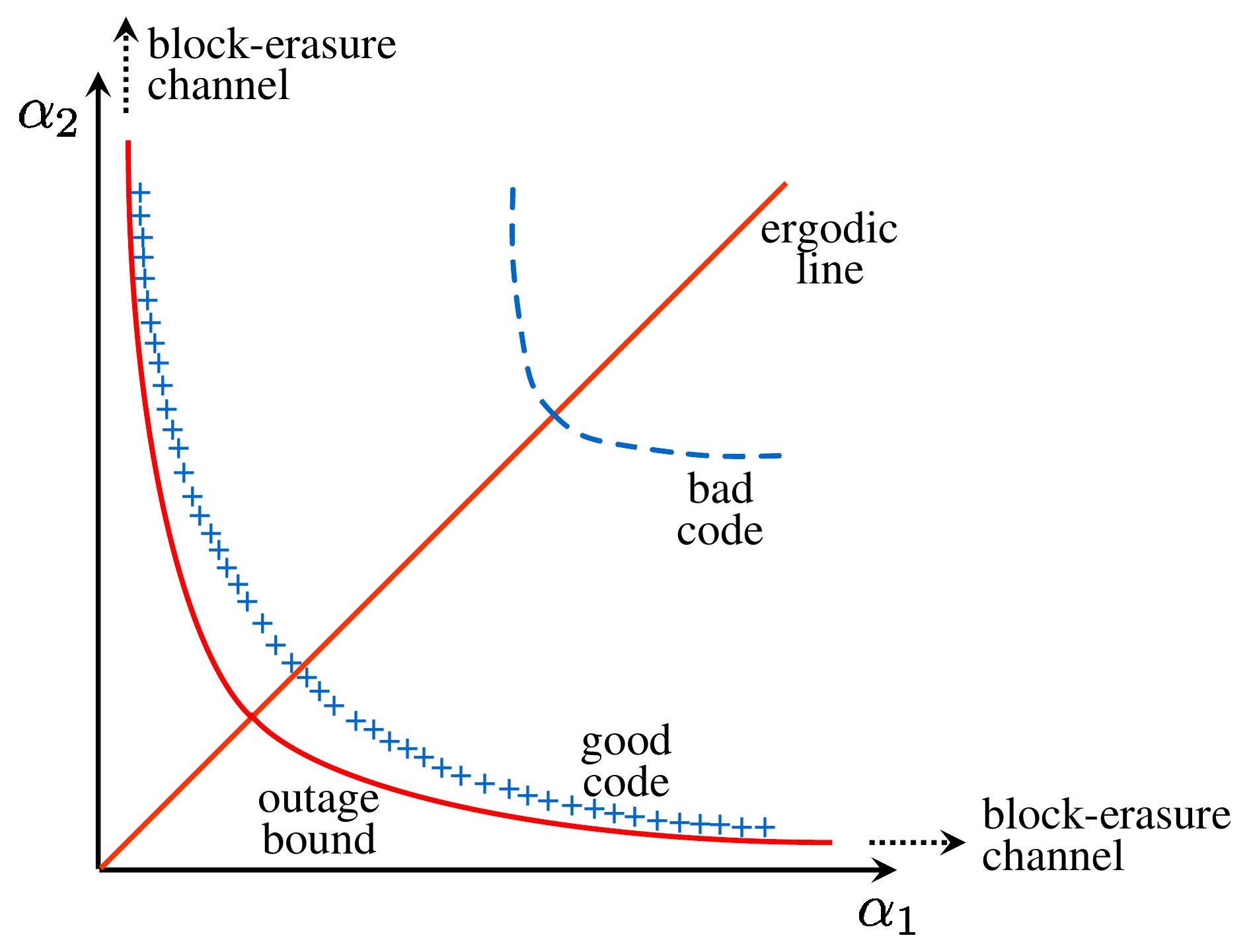}
\end{center}
\caption{\sl Outage boundaries in the fading plane for a BF
channel with $n_c=2$. To approach the outage limit, one should:
(a)~Reduce the gap on the ergodic line, which requires an
excellent decoding threshold, and (b)~Reduce the gap at infinity,
which requires a full-diversity code (MDS) on a block-erasure
channel.} \label{fig_fading_region_nc=2}
\end{figure}

We illustrate our solution to the design problem by referring to a
limiting case. Specifically, observe that, if the fading
coefficients $\alpha_i$ belong to the set $\{ 0, +\infty \}$, the
BF channel becomes a block-erasure channel \cite{Lapidoth1994,Guillen2006-c}. This corresponds to the large SNR regime. The reader is referred
to Fig.~\ref{fig_fading_region_nc=2}, where the outage boundaries
are illustrated (see~\cite{Allerton2005} for more details).

\begin{figure}[htb]
\begin{center}
\includegraphics[width=0.5\columnwidth]{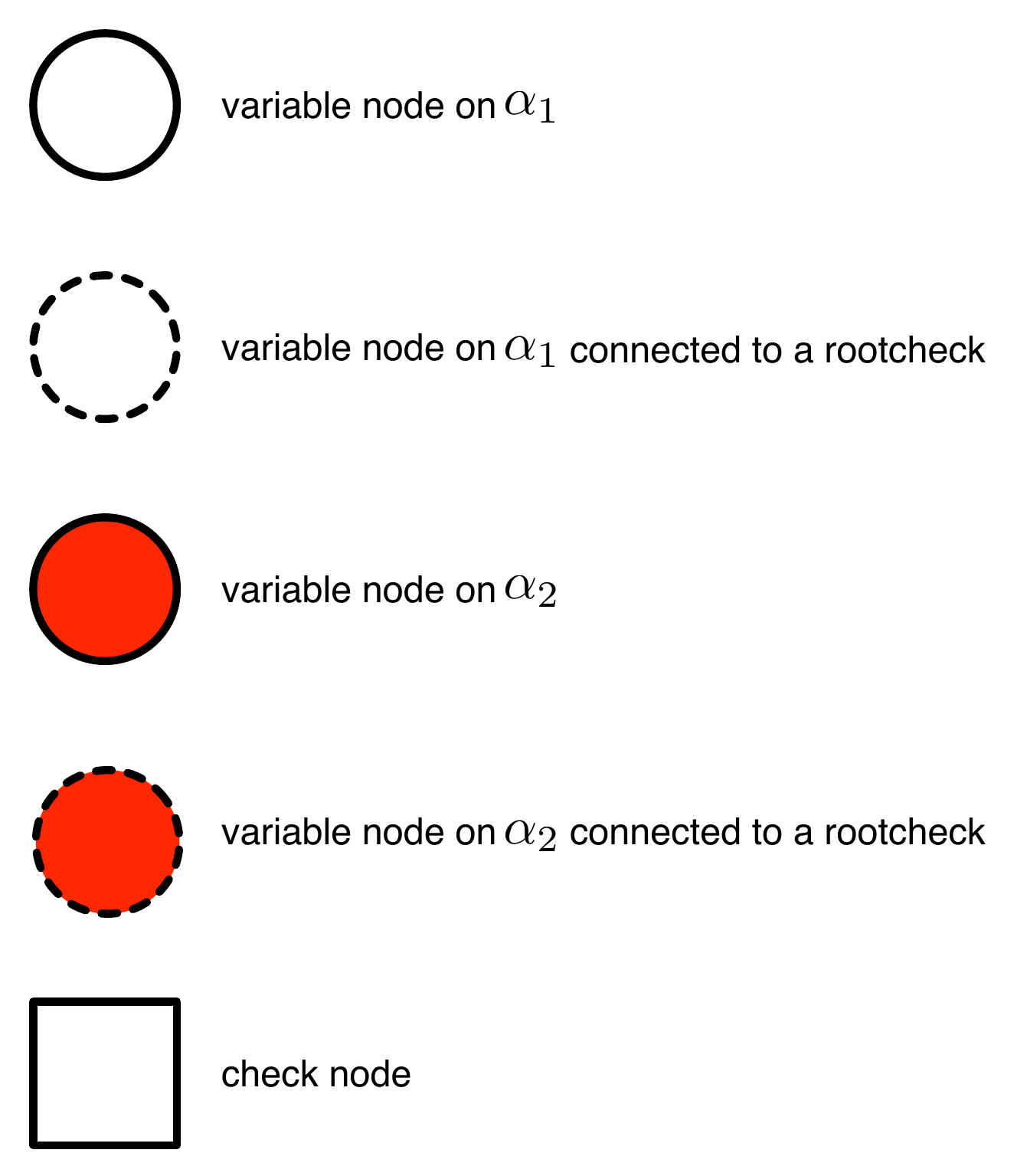}
\end{center}
\vspace{-5mm}
\caption{\sl Notations for graph representation.}
\label{fig_nodes}
\end{figure}

\begin{figure}[htb]
\begin{center}
\includegraphics[width=0.65\columnwidth]{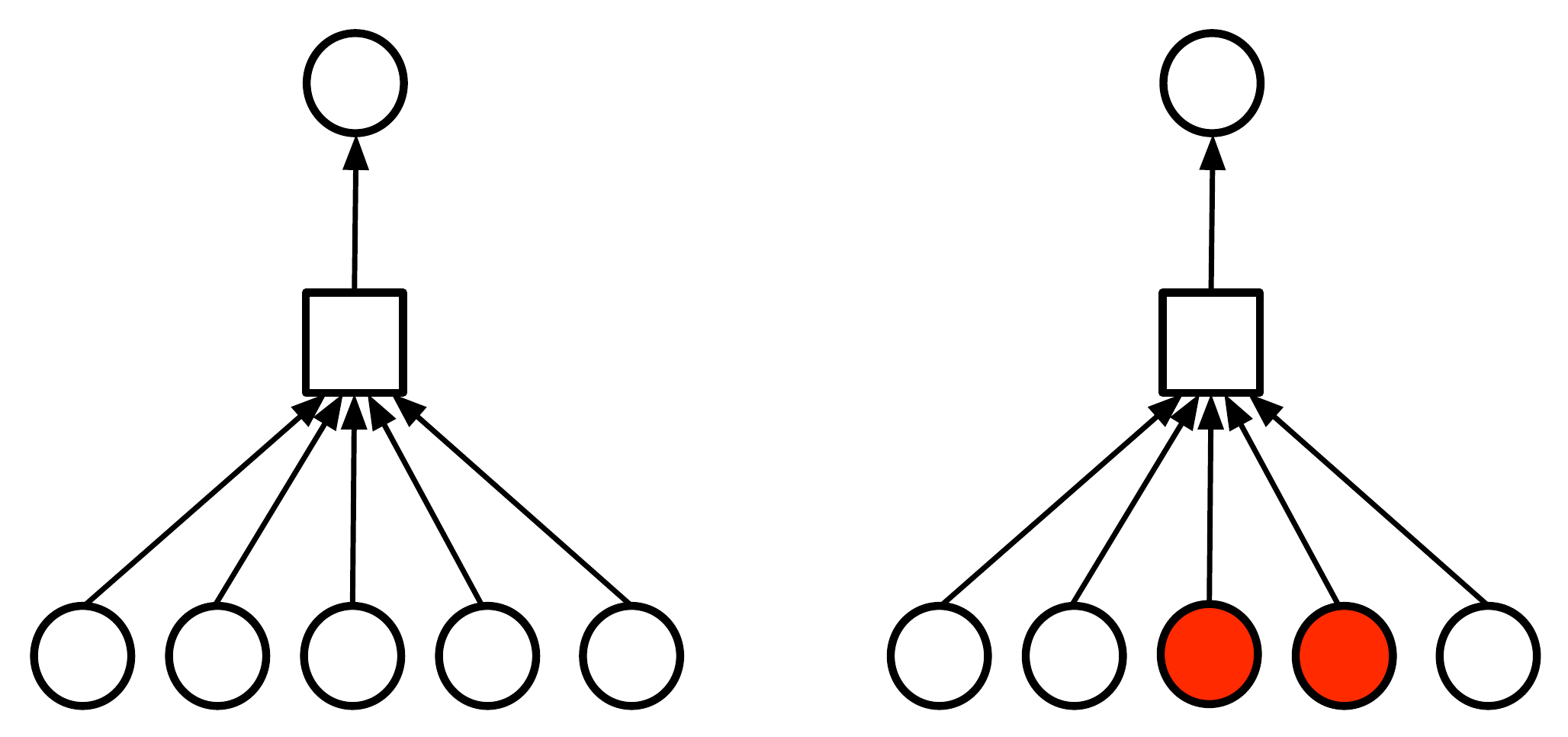}
\end{center}
\vspace{-5mm}\caption{\sl Two examples of bad configurations under belief
propagation decoding on a block-fading channel.}
\label{fig_badconfigs}
\end{figure}

In our approach, we need to find a graph whose topology yields
full diversity. For simplicity, we illustrate the case of the
$(3,6)$ LDPC ensemble with $n_c=2$ (generalizations to other
degree distributions and rates will be treated {\em infra}).
Fig.~\ref{fig_nodes} shows the notation employed in this section.
Two examples of local graphs whose diversity is not guaranteed are
shown in Fig.~\ref{fig_badconfigs}. The checknodes defining an
LDPC code are single-parity check codes, and hence they cannot
tolerate more than one erased bit. For example, if $\alpha_1=0$
then the checknodes in Fig.~\ref{fig_badconfigs} are not able to
recover the erased bit, because  it is connected to bitnodes
which are also erased, because they are subject to the same fading
coefficient. Notice also that the design must be symmetric, i.e.,
any analysis with respect to $\alpha_1$ is valid for $\alpha_2$,
and hence permuting the order of the two fading gains should yield
an equivalent design.

\begin{figure}[htb]
\begin{center}
\includegraphics[width=0.65\columnwidth]{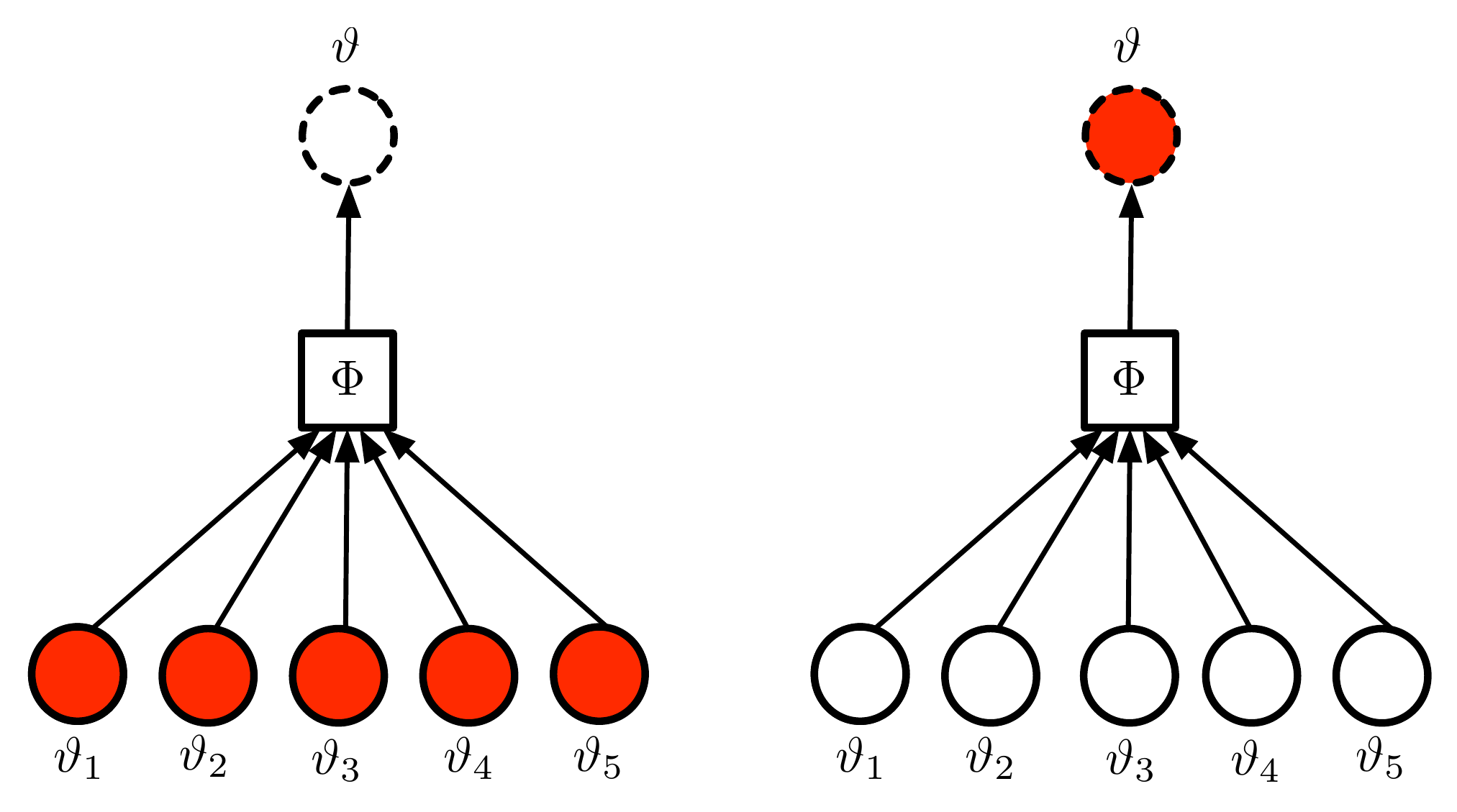}
\end{center}
\vspace{-5mm}\caption{The two unique good configurations (rootchecks)
under belief propagation decoding on a block-fading channel.}
\label{fig_goodconfigs}
\end{figure}

The two unique local graphs that guarantee full diversity in the
presence of block erasures are exhibited in
Fig.~\ref{fig_goodconfigs}. The immediate consequence is the
definition of {\em rootchecks}. We start by building a regular
$(3,6)$ structure where bitnodes have degree $3$ and checknodes
have degree $6$, next we generalize to any $(\lambda(x), \rho(x))$
degree distribution~\cite{Richardson2001-b}. A checknode $\Phi$
connected to bits $\vartheta_1, \vartheta_2, \ldots, \vartheta_6$
is written as $\Phi(\vartheta_1, \vartheta_2, \ldots,
\vartheta_6)$.

\begin{definition}
\label{def_rootcheck} Let $\vartheta$ be a binary element
transmitted on fading $\alpha_1$. A type-$1$ rootcheck for
$\vartheta$ is a checknode $\Phi(\vartheta, \vartheta_1, \ldots,
\vartheta_5)$ where all bits $\vartheta_1, \ldots, \vartheta_5$
are transmitted on fading $\alpha_2$.
\end{definition}

\vspace{12pt} Type-$2$ rootchecks are defined similarly.

\begin{figure}[htb]
\begin{center}
\subfigure[\label{fig_tannergraph_rootldpc}Tanner graph.]{\includegraphics[width=0.6\columnwidth]{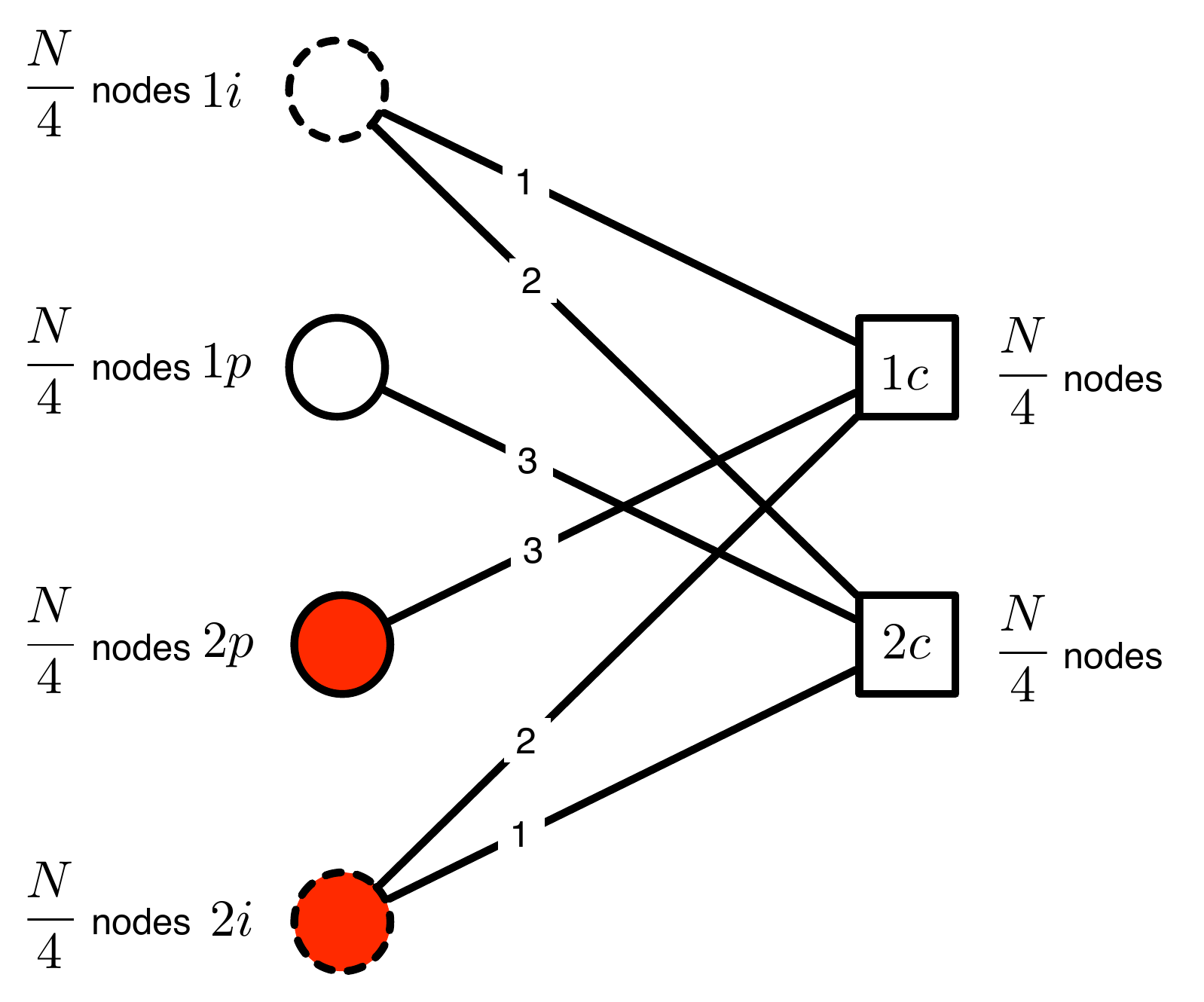}}
\subfigure[\label{fig_matrix_rootldpc}Parity-check matrix.]{\includegraphics[width=0.7\columnwidth]{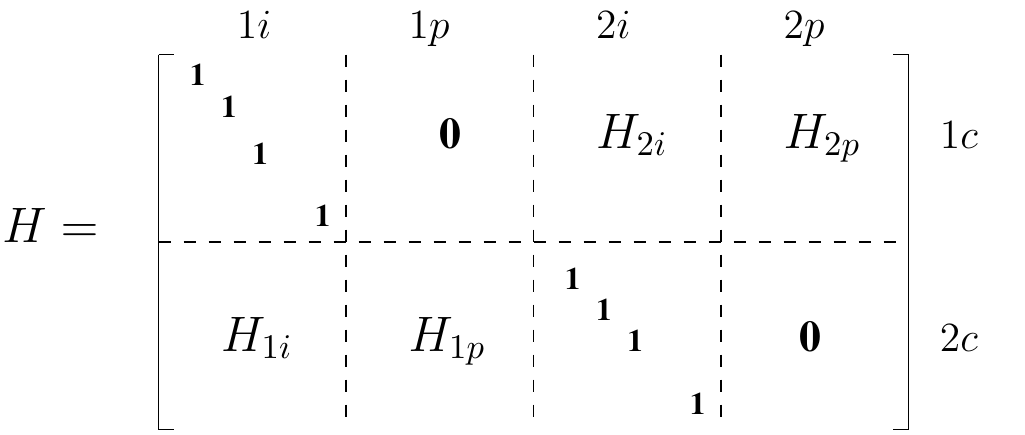}}
\end{center}
\caption{Tanner graph and parity-check matrix for a regular (3,6) root-LDPC code of rate $1/2$.
An irregular structure $(\lambda(x),\rho(x))$ can be easily plugged on edges connected
to non-root checknodes.}
\end{figure}

\begin{figure}[htb]
\begin{center}
\includegraphics[width=0.7\columnwidth]{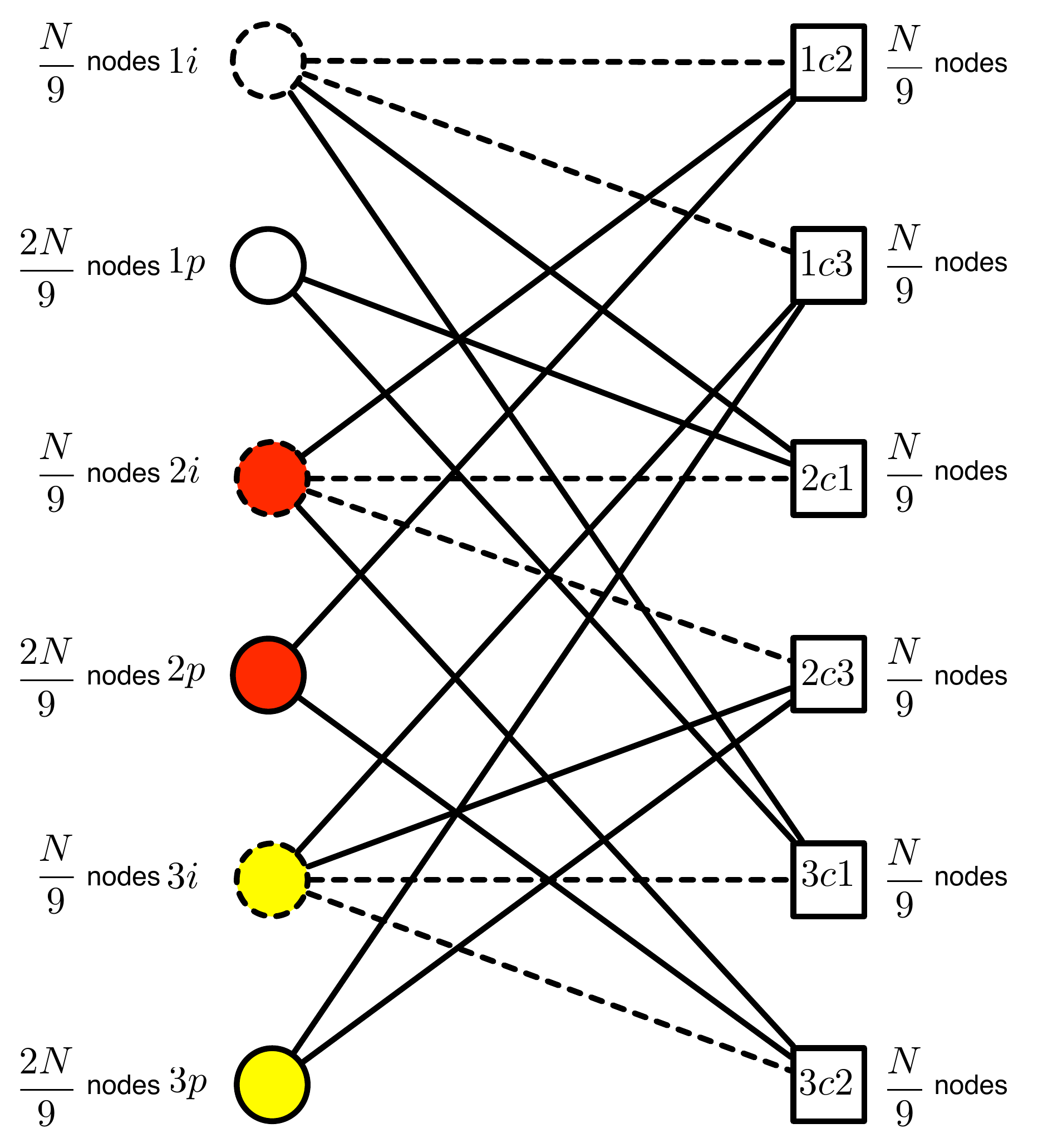}
\end{center}
\caption{Tanner graph for a regular (4,6) root-LDPC code of rate $1/3$.
The introduction of any $(\lambda(x),\rho(x))$ irregularity is always possible
on edges connected to non-root checknodes.}
\label{fig_tannergraph_rootldpc_1_3}
\end{figure}

 Using Definition~\ref{def_rootcheck}, consider a
length-$N$, rate-$1/2$ LDPC code. Information bits are split into
two classes: $N/4$ bits (tagged $1i$) are transmitted on
$\alpha_1$, while $N/4$ bits (tagged $2i$) 
are transmitted on
$\alpha_2$. Parity bits are also partitioned into two sets, say
$1p$ and $2p$. Finally, we connect all information bits to
rootchecks in order to guarantee full diversity when word error
probability is measured on those bits. The protection of parity
bits is abandoned. This design produces the bipartite Tanner graph
drawn in Fig.~\ref{fig_tannergraph_rootldpc}. Its extension to
rate $1/3$ is portrayed in
Fig.~\ref{fig_tannergraph_rootldpc_1_3}. Integers labeling edges
indicate the degree of a node along those edges. The structure of
$H$ for a root-LDPC code is directly derived from its Tanner
graph, and is shown in Fig.~\ref{fig_matrix_rootldpc}. The $N/4
\times N/4$ identity matrix is written twice in connections $1i
\leftrightarrow 1c$ and $2i \leftrightarrow 2c$. Two all-zero $N/4
\times N/4$ submatrices prohibit any edge of type $1p
\leftrightarrow 1c$ and $2p \leftrightarrow 2c$. The other $4$
submatrices are all sparse, $H_{1i}$ and $H_{2i}$ are random
sparse matrices of Hamming weight $2$ per row and per column.
Similarly, $H_{1p}$ and $H_{2p}$ are random sparse matrices of
Hamming weight $3$ per row and per column.

An irregular version of a root-LDPC code can be built from a left degree distribution
$\lambda(x)$ and a right degree distribution $\rho(x)$ by appropriately modifying the weight
distribution of the $4$ sub-matrices $H_{1i}$, $H_{2i}$, $H_{1p}$, and $H_{2p}$.
Equivalently, the degree distribution changes the distribution of edges connected
to non-rootchecks in the Tanner graph. Irregularity has no influence on the diversity order
because rootchecks are maintained. Irregularity should enhance the coding gain
by pushing the code boundary near the outage capacity limit on the ergodic line.

\begin{proposition}
\label{prop_fulldiv_erasure} Consider a rate-$R=1/2$ root-LDPC
code with degree distribution $(\lambda(x), \rho(x))$ transmitted
on a block-erasure channel with $n_c=2$. Then, under iterative
message passing decoding, the root-LDPC code has full-diversity.
\end{proposition}
\begin{proof}
The two fading coefficients  $\alpha_1$ and $\alpha_2$ are independent
and take two possible values $\{0, +\infty \}$.
Examining the  Tanner graph of
Fig.~\ref{fig_tannergraph_rootldpc}, we observe that the only outage
event occurs when $\alpha_1=\alpha_2=0$ (both blocks erased).
Indeed, when $\alpha_1=0$ and $\alpha_2=+\infty$, it is
straightforward to see that  information bits $1i$  are determined
using rootchecks $1c$. Similarly,  when $\alpha_1=+\infty$ and
$\alpha_2=0$, information bits $2i$  are determined  using
rootchecks $2c$.
\end{proof}
\vspace{12pt}

On a block-erasure channel, let $\epsilon$ be the probability that
$\alpha_i$ be equal to $0$. From the proof of
Proposition~\ref{prop_fulldiv_erasure} above, we find that the
word error probability of a root-LDPC code is $\epsilon^2$. As shown
in \cite{Guillen2006-c}, this is precisely the outage probability of
the channel, and therefore, full-diversity blockwise MDS codes are outage
achieving in the block-erasure channel. As remarked in
\cite{Guillen2006-c}, blockwise MDS codes are necessary, but not sufficient to
achieve the outage limit in noisy channels. In the following, we study
the behavior of root-LDPC over general Rayleigh BF AWGN channels.

\subsection{The general case}
Now we study the general case of Rayleigh BF. Some simple facts
about $4$th-order $\chi^2$ distributions are reviewed in the
Appendix. In the sequel, we use the notations of the Appendix to
analyze the diversity metric in log-ratio messages.

\begin{proposition}
\label{prop_fulldiv_rayleigh} Consider a rate-$1/2$,
$(\lambda(x), \rho(x))$ root-LDPC code transmitted on a 
Rayleigh block-fading channel with $n_c=2$. Then, under iterative belief
propagation decoding, the root-LDPC code has full-diversity.
\end{proposition}
\begin{proof}
As indicated in the design of a root-LDPC code before
Proposition~\ref{prop_fulldiv_erasure}, the diversity order of a
root-LDPC code does not depend on its left or right degree
distribution. This can also be proved via the evolution trees in
the next section. Thus, we restrict this proof to a regular
$(3,6)$ LDPC. The extension to the irregular case is straighforward.

Let $\Lambda^a_i$, $i=1 \ldots  \delta-1$, denote the input
log-ratio probabilistic messages to a checknode $\Phi$ of degree
$\delta$. The output message $\Lambda^e$ for belief propagation is
\begin{equation}
\Lambda^e = 2 ~\tthh^{-1} \left(\prod_{i=1}^{\delta-1} \tthh\left(\frac{\Lambda^a_i}{2}\right) \right)
\end{equation}
where $\tthh(x)$ denotes the hyperbolic-tangent function.
Superscripts $a$ and $e$ stand for {\em a priori} and {\em
extrinsic}, respectively. In order to simplify the proof, we will
show that a suboptimal belief propagation decoder is able to
achieve diversity order $2$. Therefore, if a suboptimal decoder
achieves full diversity, the optimal decoder also achieves full
diversity. Consider the min--sum decoder. The output message
produced by a checknode $\Phi$ is now approximated by
\begin{equation}
\label{equ_approx_L}
\Lambda^e=\min(|\Lambda^a_i|) \prod_{i=1}^{\delta-1} \sign(\Lambda^a_i)
\end{equation}
\paragraph{First decoding iteration}
We first study the output after one decoding iteration. We assume that the all-zero codeword has been transmitted. The channel
crossover probability associated with fading $\alpha_j$, $j=1,2$,
is
\[
\epsilon_j=Q \left( \sqrt{ 2\gamma\alpha_j^2} \right)
\]
The channel message for a bit $\vartheta$ transmitted over fading coefficient $\alpha$ is
\begin{equation}
\label{equ_L0}
\Lambda_0 = \log \left( \frac{p(y|\vartheta=0, \alpha)}{p(y|\vartheta=1, \alpha)} \right) = \frac{2\alpha y}{\sigma^2}
~=~ \frac{2}{\sigma^2} (\alpha^2+\alpha z)
\end{equation}
where $y=\alpha+z$ and $z \sim \mathcal{N}(0,\sigma^2)$ (assuming $\Es=1$).
At the first decoding iteration, all input messages $\Lambda^a_i$ in (\ref{equ_approx_L})
have an expression identical to (\ref{equ_L0}).

An information bit $\vartheta$ of class $1i$ has $\Lambda_0=\frac{2}{\sigma^2} (\alpha_1^2+\alpha_1 z_0)$.
It also receives 3 messages $\Lambda^e_i$, $i=1 \ldots 3$ from its 3 neighboring checknodes.
The total {\em a posteriori} message corresponding to $\vartheta$ is $\Lambda=\Lambda_0+\Lambda^e_1+\Lambda^e_2+\Lambda^e_3$.
Let $\Lambda^e_1$ be the extrinsic message generated by the rootcheck of class $1c$ connected to $\vartheta$.
The error rate $P_e(1i)$ on class $1i$ is given by the negative tail of the density of $\Lambda$ messages.
The addition of $\Lambda^e_2+\Lambda^e_3$ to $\Lambda_0+\Lambda^e_1$ cannot degrade $P_e(1i)$
because the convolution with the density of messages from non-rootchecks can only physically upgrade
the resulting density. Thus, it is sufficient to prove that message $\Lambda_0+\Lambda^e_1$ brings full diversity.
The expression of $\Lambda^e_1$ is found by applying (\ref{equ_approx_L}). Input messages to the rootcheck
are negative with probability $\epsilon_2$. Then
\[
\Lambda^e_1= S_1 \frac{2}{\sigma^2} (\alpha_2^2+\alpha_2 z_1)
\]
where
\[
S_1 = \sum_{i~\text{even}} \binom{4}{i} \epsilon_2^i
(1-\epsilon_2)^{4-i} - \sum_{i~\text{odd}}\binom{4}{i} \epsilon_2^i
(1-\epsilon_2)^{4-i}
\]

We obtain
\[
\Lambda^e_1 = (1-2\epsilon_2)^4 \frac{2}{\sigma^2} (\alpha_2^2+\alpha_2 z_1)
\]
The partial {\em a posteriori} log-ratio message becomes
\[
\Lambda_0+\Lambda^e_1=
\frac{2}{\sigma^2} \left( \alpha_1^2+(1-2\epsilon_2)^4\alpha_2^2\right) ~+~\alpha_1 z_0 + (1-2\epsilon_2)^4 \alpha_2 z_1)
\]
The embedded metric $Y=\alpha_1^2+(1-2\epsilon_2)^4 ~\alpha_2^2$
guarantees full diversity. At high SNR (i.e., when $E_b/N_0
\rightarrow +\infty$), $Y$ behaves exactly as
$\alpha_1^2+\alpha_2^2$.

\paragraph{Further decoding iterations}
As can be seen from the decoding tree of a bitnode $1i$ in Fig.
\ref{fig_DE_tree_2}, the diversity order 2 is maintained after the
first iteration. Indeed, at the input of the rootcheck,
information bits of class $2i$ have already full diversity and
parity bits $2p$ bring always a term proportional to $\alpha_2^2$.
Due to the particular structure of root-LDPC codes, the density of message $\Lambda_0+\Lambda^e_1$ can only be
improved with respect to the first iteration. Hence, full
diversity is preserved.
\end{proof}
\vspace{12pt}

 The proof of the previous proposition is based on
showing that the information bits have diversity $2$. In the
following, we examine the diversity of the parity bits.
A parity bit $\vartheta$ of class $1p$ has $\Lambda_0=\frac{2}{\sigma^2} (\alpha_1^2+\alpha_1 z_0)$.
It also receives 3 messages $\Lambda^e_i$, $i=1 \ldots 3$ from its 3 neighboring checknodes
all of class $2c$. The total {\em a posteriori} message of $\vartheta$ is $\Lambda=\Lambda_0+\Lambda^e_1+\Lambda^e_2+\Lambda^e_3$.
Now let us determine the nature of $\Lambda^e_i$ based on input messages to a checknode $\Phi$
of class $2c$ as illustrated in Figures \ref{fig_tannergraph_rootldpc} and \ref{fig_DE_tree_3}.
The node $\Phi$ is not a rootcheck. We need to determine the metric $Y$ embedded in its output message.
In the case $\alpha_2 \le \alpha_1$ (this happens with probability $1/2$), it can be shown
that, after one decoding iteration, the extrinsic message produced by $\Phi$ satsifies
%
\begin{equation}
\Lambda^e_i =
\begin{cases}
S\frac{2}{\sigma^2} (\alpha_2^2+\alpha_2 z) &\text{with probability $G^4\ge\frac{1}{16}$}\\
S\frac{2}{\sigma^2} (\alpha_1^2+\alpha_1 z) &\text{with probability $1-G^4\le\frac{15}{16}$}
\end{cases}\nonumber
\end{equation}
where the function $G$ is defined in the Appendix. On the
contrary, when $\alpha_2 \ge \alpha_1$, it can be shown that
\begin{equation}
\Lambda^e_i =
\begin{cases}
S\frac{2}{\sigma^2} (\alpha_2^2+\alpha_2 z) &\text{with probability $G^4\le\frac{1}{16}$}\\
S\frac{2}{\sigma^2} (\alpha_1^2+\alpha_1 z) &\text{with probability $1-G^4\ge\frac{15}{16}$}
\end{cases}\nonumber
\end{equation}
%
We conclude that, for parity bits, with a probability greater than
$\frac{1}{2} \times \frac{15}{16}$, the output message has
diversity order one. In spite of the presence (with a nonzero
probability) of diversity-$2$ messages, the error probability of
parity bits will be dominated by weak messages with diversity $1$.
The above arguments are still valid for further decoding
iterations.

Finally, we look at the minimum partial Hamming weight $\omega^\star$ under belief propagation decoding.
\begin{corollary}
A root-LDPC code with $R=1/2$ has full-diversity under iterative belief propagation decoding.
\end{corollary}
\begin{proof}
Consider an  information bit $\vartheta$ of class  $1i$. Let $\delta_b
\ge  2$ be  the degree  of $\vartheta$.   At high  SNR,  the log-ratio
message   produced   by  its   rootcheck   has   an  embedded   metric
$\alpha_1^2+\alpha_2^2$.   Consider  the  $\delta_b-1$ non-root  checknodes
connected to  $\vartheta$.   Since parity  bits  of class  $1p$
dominate the error  probability at the input of  $2c$ checknodes, then
its  metric will  be  $\alpha_1^2$. Finally,  the  {\em a  posteriori}
log-ratio  message associated  to  $\vartheta$ will  contain a  metric
of the type $\delta_b     \alpha_1^2+\alpha_2^2$ which has diversity 2.
\end{proof}
\vspace{12pt}

\begin{figure}[t!]
\begin{center}
\includegraphics[width=0.99\columnwidth]{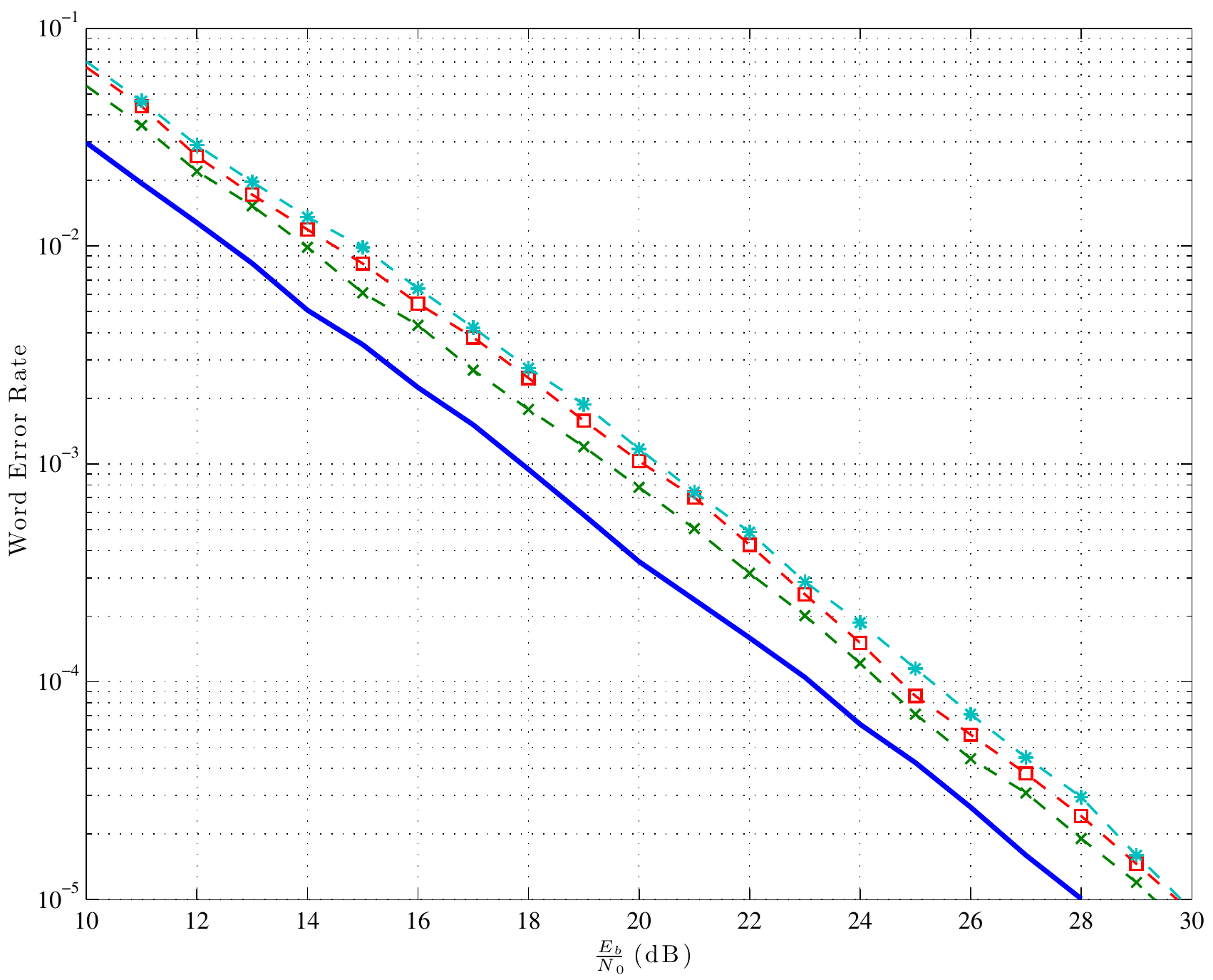}
\end{center}
\caption{\sl Regular (3,6) root-LDPC codes with iterative decoding
on a Rayleigh block-fading channel with $n_c=2$. Word-error rate
is measured on information bits. The thick solid line corresponds to the outage probability with BPSK, the dotted lines with $\times$ markers correspond to $N=200$, the dotted lines with $\square$ markers correspond to $N=2000$ and the dotted lines with markers $*$ correspond to $N=20000$.} \label{fig_perf_ldpc_root}
\end{figure}

In Fig.~\ref{fig_perf_ldpc_root}, we  illustrate the performance
of the $(3,6)$ root-LDPC ensemble. As  we observe, the performance
is similar for all  ranges of  $N$, and  it is also close to the
outage probability  of the channel.  This  effect was first
observed with blockwise-concatenated codes and repeat-accumulate codes in \cite{Guillen2006-a}, and then in
\cite{Allerton2004,Allerton2005,ITA2006} for 
parallel turbo codes. This effect is due to the threshold behavior of
{\em good} codes, i.e., for a given channel realization, the code has
a SNR threshold (independent of $N$) below which the decoder cannot
decode successfully. Hence, whenever this threshold is larger than the
SNR $\gamma$, the decoder will 
make an error
for sufficiently large word length \cite{Guillen2006-a}. This is considered in more detail in the following section, where the analysis of the word error probability under iterative decoding for large $N$ is done using density evolution.

%
\section{Density evolution in presence of block fading}\label{sectionV}
\begin{figure}[htb]
\begin{center}
\includegraphics[width=0.55\columnwidth]{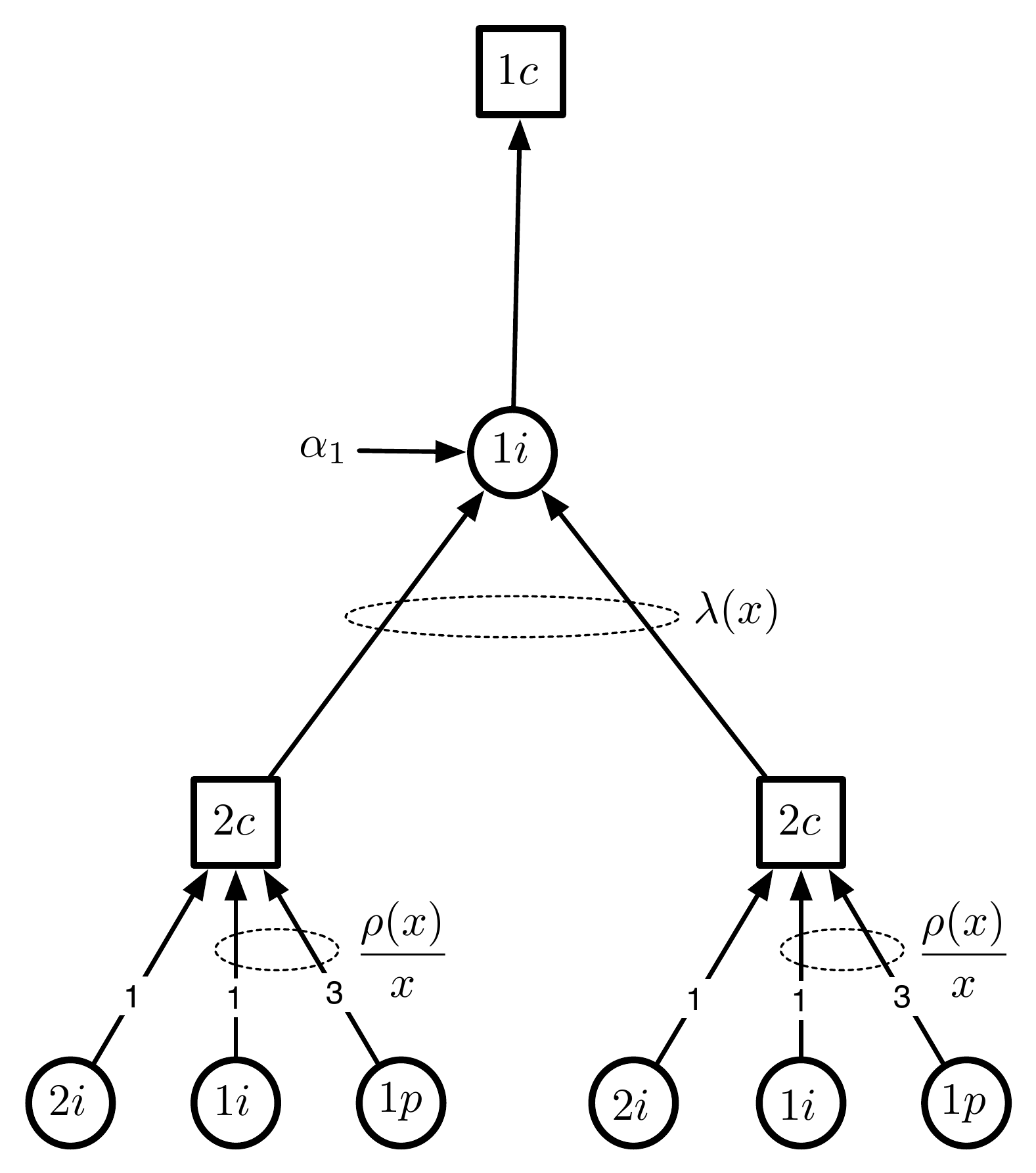}
\end{center}
\caption{\sl Local neighborhood of bitnode $1i$. This tree is used
to determine the evolution of messages $1i \rightarrow 1c$.}
\label{fig_DE_tree_1}
\end{figure}

\begin{figure}[htb]
\begin{center}
\includegraphics[width=0.55\columnwidth]{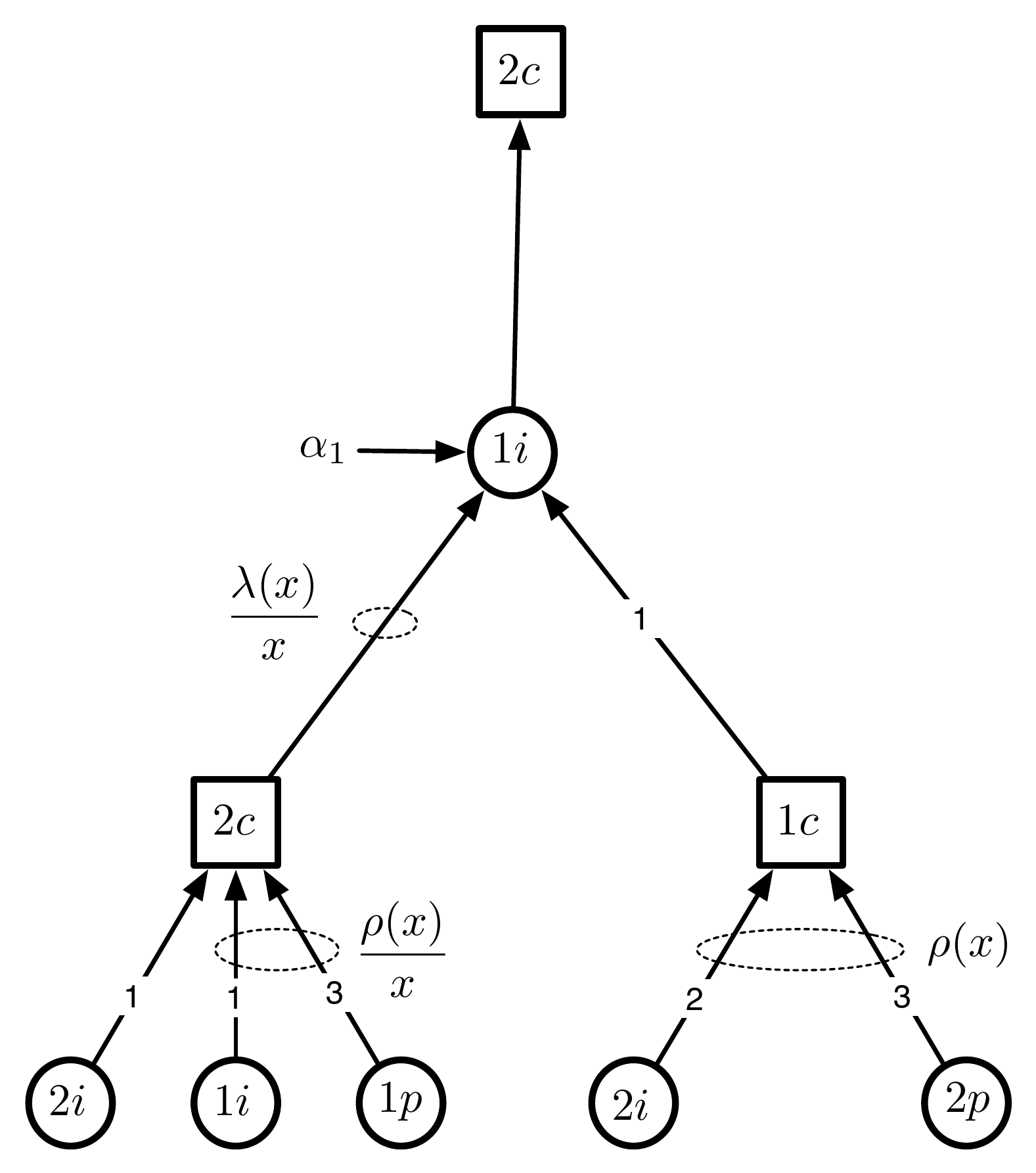}
\end{center}
\caption{\sl Local neighborhood of bitnode $1i$. This tree is used
to determine the evolution of messages $1i \rightarrow 2c$.}
\label{fig_DE_tree_2}
\end{figure}

\begin{figure}[htb]
\begin{center}
\includegraphics[width=0.55\columnwidth]{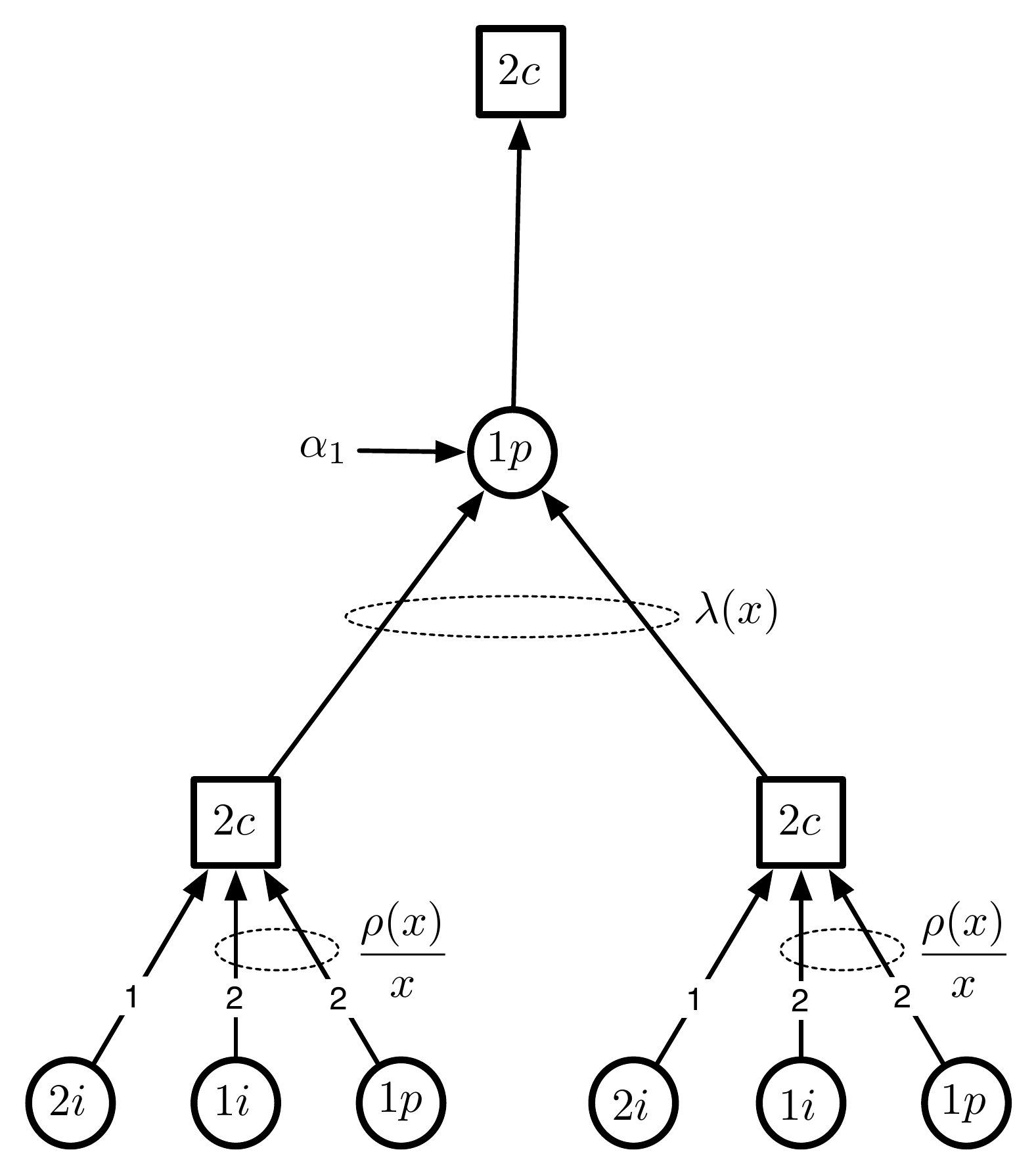}
\end{center}
\caption{\sl Local neighborhood of bitnode $1p$. This tree is used
to determine the evolution of messages $1p \rightarrow 2c$.}
\label{fig_DE_tree_3}
\end{figure}

 The evolution of message
densities~\cite{Richardson2001-a,Richardson2007}  under iterative
decoding is  described  through  six evolution  trees for  a
binary LDPC root-code.  The evolution trees represent the local
neighborhood of a bitnode in an infinite-length code whose graph
has no cycles. Figs.~\ref{fig_DE_tree_1}, \ref{fig_DE_tree_2}, and
\ref{fig_DE_tree_3} show the local neighborhoods of classes $1i$
and  $1p$.
 Similar evolution trees can be  drawn for  classes $2i$
and  $2p$.  Full diversity in  the presence of fading is
guaranteed, thanks to messages  $1c \rightarrow 1i$ (respectively,
$2c \rightarrow 2i$)  as indicated in the  proof of
Proposition~\ref{prop_fulldiv_rayleigh}. Irregularity is defined
in the standard way~\cite{Richardson2001-b} through the
polynomials $\lambda(x)$ and $\rho(x)$. The polynomial
$\lambda(x)$ is replaced by $\tilde{\lambda}(x)=\lambda(x)/x$ each
time an edge is isolated at the input of  a bitnode.   In
addition, the polynomial $\rho(x)$ is  replaced by
$\tilde{\rho}(x)=\rho(x)/x$ each time an edge is isolated at the
input of a checknode.  The following notations are used, where the
superscript $m$ is an integer denoting the decoding iteration
order:
\begin{itemize}
\item $q^m_1(x)$ and $q^m_2(x)$: Probability density functions of log-ratio messages
on the edges $1i \rightarrow 1c$ and $2i \rightarrow 2c$
respectively. See Fig. \ref{fig_DE_tree_1}.
\item $f^m_1(x)$ and $f^m_2(x)$: Probability density functions of log-ratio messages
on the edges $1i \rightarrow 2c$ and $2i \rightarrow 1c$
respectively. See Fig. \ref{fig_DE_tree_2}.
\item $g^m_1(x)$ and $g^m_2(x)$: Probability density functions of log-ratio messages
on the edges $1p \rightarrow 2c$ and $2p \rightarrow 1c$
respectively. See Fig. \ref{fig_DE_tree_3}.
\item Let $X_1 \sim p_1(x)$ and $X_2 \sim p_2(x)$ be two independent real random variables.
The density function of $X_1+X_2$ obtained by convolving the two original densities
is written as $p_1(x) \otimes p_2(x)$. The notation $p(x)^{\otimes n}$ denotes the convolution
of $p(x)$ with itself $n$ times. The expression $\lambda(p(x))$ represents the density
function $\sum_i ~\lambda_i ~p(x)^{\otimes i-1}$.
\item Let $X_1 \sim p_1(x)$ and $X_2 \sim p_2(x)$ be two independent real random variables.
The density function $p(y)$ of the variable $Y=2
\tthh^{-1}(\tthh(\frac{X_1}{2})\tthh(\frac{X_2}{2}))$ obtained
through a checknode is written as $p_1(x) \odot p_2(x)$ and is
called {\em R-convolution} \cite{Richardson2007}. The notation $p(x)^{\odot n}$ denotes
the R-convolution of $p(x)$ with itself $n$ times. The expression
$\rho(p(x))$ represents the density function $\sum_i ~\rho_i
~p(x)^{\odot i-1}$.
\end{itemize}

\begin{proposition}
Consider an (ergodic) AWGN channel (i.e., assume
$\alpha_1=\alpha_2=1$). Under iterative decoding, a $(\lambda(x),
\rho(x))$ root-LDPC code has the same decoding threshold as a
random $(\lambda(x), \rho(x))$ LDPC code.
\end{proposition}
\begin{proof}
With the two fading gains equal to unity, the six evolution trees
degenerate into a single tree, and all densities become identical:
$q^m_1(x)=q^m_2(x)=f^m_1(x)=f^m_2(x)=g^m_1(x)=g^m_2(x)$ for any
decoding iteration $m$. Thus, density evolution of a root-LDPC
code reduces to a classical density evolution of a random code
given by $p^{m+1}(x)=\lambda(\rho(p^m(x)))$.
\end{proof}
\vspace{12pt}

\begin{proposition}
\label{equ_DE_equations} Consider a nonergodic BF channel with
$n_c=2$. For fixed fading coefficients $(\alpha_1, \alpha_2)$, the
density evolution equations of a $(\lambda(x), \rho(x))$ root-LDPC
code are, for all $m$,
\[
\begin{array}{lll}
q_1^{m+1}(x) &=& \mu_1(x)  \otimes~ \lambda \left( q_2^m(x) \odot
\tilde{\rho} \left( f_e ~f_1^m(x) + g_e ~g_1^m(x)
\right) \right)\\
f_1^{m+1}(x) &=& \mu_1(x) \otimes~ \tilde{\lambda} \left( q_2^m(x)
\odot \tilde{\rho} \left( f_e ~f_1^m(x) + g_e ~g_1^m(x) \right)
\right) \otimes~ \rho \left( f_e ~f_1^m(x) + g_e ~g_1^m(x) \right)\\
g_1^{m}(x) &=& q_1^{m}(x)
\end{array}
\]
where the multi-edge type fraction is
\[
f_e=1-g_e=\frac{\sum_i \lambda_i/i}{\sum_i  \lambda_i/(i-1) +
\sum_i \lambda_i/i}
\]
and $\mu_1(x)$ is the Gaussian density at the output of the
channel with fading $\alpha_1$. Similar density evolution
equations are obtained by permuting the two fading gains.
\end{proposition}
\begin{proof}
The above equations are directly derived from local neighborhoods
of bitnodes in the graphical representation of the LDPC code, following standard density evolution analysis of multi-edge type LDPC codes \cite{Richardson2007}.
\end{proof}
\vspace{12pt}

\begin{figure}[htb]
\begin{center}
\includegraphics[width=0.99\columnwidth]{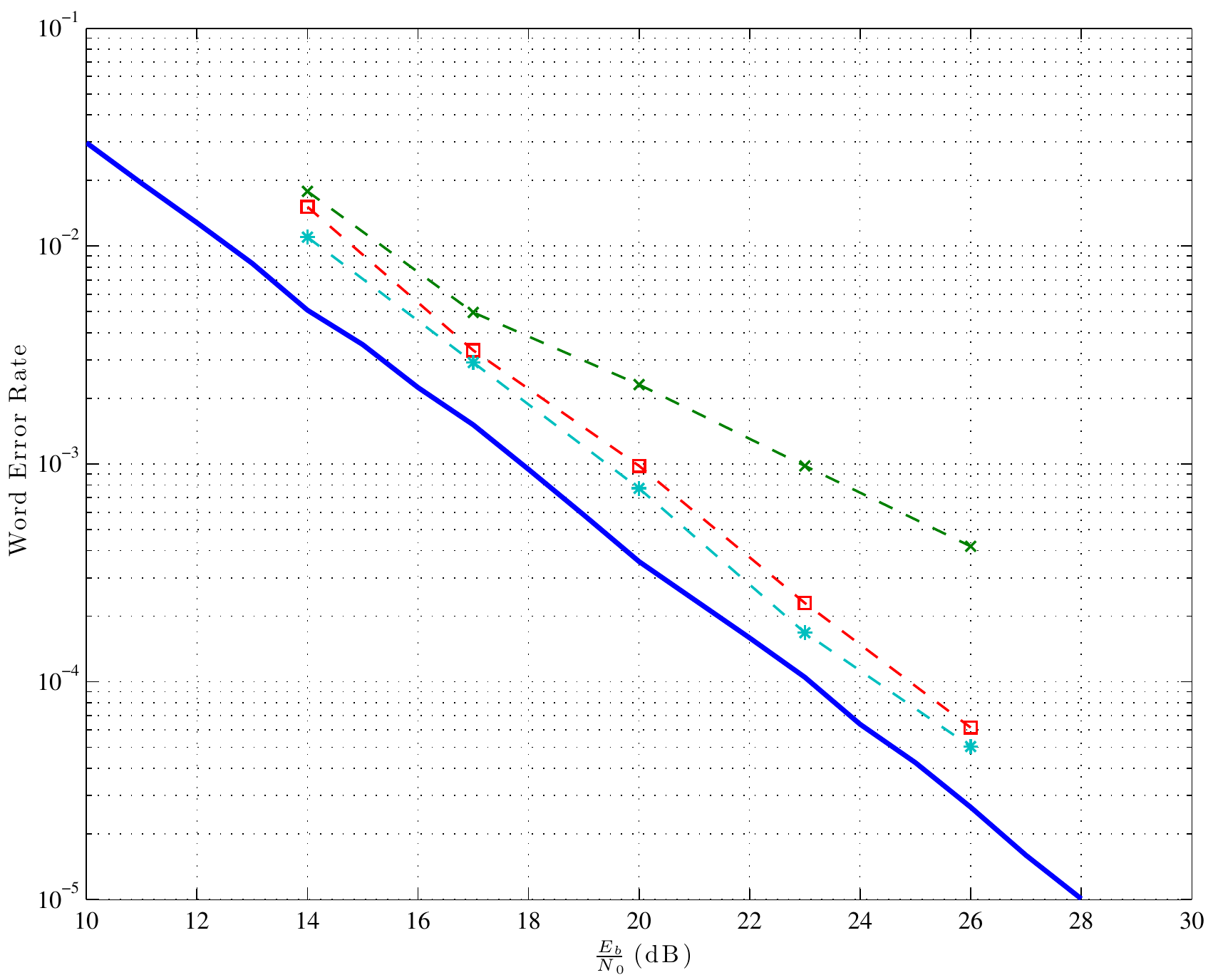}
\end{center}
\caption{\sl Density evolution of random-LDPC and root-LDPC codes
with iterative decoding on a block-fading channel with $n_c=2$. The thick solid lines correspond to the outage probability with BPSK, the  dotted lines with $\times$ markers correspond to the random LDPC, the dotted lines with $\square$ markers correspond to the $(3,6)$ root LDPC and the  dotted lines with $*$ markers correspond to the irregular root LDPC ensemble.}
\label{fig_perf_ldpc_DE}
\vspace{-5mm}
\end{figure}

To evaluate the performance of  LDPC codes via density evolution
in presence  of  nonergodic  fading,  we  illustrate the results
obtained by applying Proposition~\ref{equ_DE_equations} to the
calculation of asymptotic error probability of the code, in a similar way to what is done in \cite{Allerton2005}. Three codes are shown in
Fig.~\ref{fig_perf_ldpc_DE}: a random $(3,6)$ regular code,  a
root $(3,6)$ regular code, and an  LDPC irregular root code with
left and right degree distributions given by the
polynomials~\cite{Richardson2001-b} 
\[
\lambda(x)=0.24426x+0.25907x^2+0.01054x^3+0.05510x^4+0.01455x^7+0.01275x^9+0.40373x^{11}
\]
and
\[
\rho(x)=0.25475x^6+0.73438x^7+0.01087x^8.
\]
\vspace{-10mm}

Refer again to the outage boundary representation in the fading
plane of~Fig.~\ref{fig_fading_region_nc=2}. Let $\alpha_0$ be the
fading value defined  by the  intersection  of the BPSK outage
boundary and  the ergodic line.  For rate $1/2$, this intersection
point satifies $I_b(\alpha^2_0 E_b/N_0)=1/2$, where
$I_b(x)\triangleq I_{\rm AWGN}(Rx)$  is the average mutual
information on an AWGN channel with a binary input and an SNR per
bit equal to $x$.

Let $\alpha_{\text{th}}$ denote the fading value defined by the
intersection of the LDPC code outage boundary and the ergodic
line. Then we have
\[\alpha_{\text{th}}^2=\frac{\left. \frac{\Eb}{N_0}\right|_{\text{th}}}{\frac{\Eb}{N_0}},\]
where $\left. \frac{E_b}{N_0}\right|_{\text{th}}$ is the decoding
threshold of the LDPC code over the ergodic AWGN channel. Finally,
we obtain
\[
\alpha_{\text{th}} ~=~ \alpha_0  \sqrt{ \frac{\left.
\frac{E_b}{N_0}\right|_{\text{th}}}{I_b^{-1}(\frac{1}{2})}} ~=~
\alpha_0 \sqrt{\Delta}
\]
where $\Delta$ in the signal-to-noise  ratio gap separating the
decoding threshold and the capacity limit on the Gaussian channel.
To better understand the gain due to irregularity illustrated in
Fig. \ref{fig_perf_ldpc_DE}, we evaluate the ratio
$\alpha_{\text{th}}/\alpha_0$.
\begin{itemize}
\item For the regular (3,6) LDPC code, the threshold is $1.09$~dB above the Gaussian channel.
Hence, $\alpha_{\text{th}}/\alpha_0=1.107$.
\item For the irregular LDPC code given above,  the threshold is $0.37$dB above the Gaussian channel.
Hence, $\alpha_{\text{th}}/\alpha_0=1.045$.
\end{itemize}

Using the best irregular code proposed in~\cite{Richardson2001-b}
with a threshold of  $0.25$~dB, we obtain
$\alpha_{\text{th}}/\alpha_0=1.007$. Hence, with
$\alpha_c/\alpha_0$ close  to $1$, the area between  the outage
capacity boundary and the code outage  boundary is decreased in
the neighborhood of the ergodic line.  However, this does not ensure
that,  the code
outage boundary  would be close to the outage capacity boundary in the  critical region
between the ergodic  line and the block-erasure  channel. Therefore, in order
to approach  the outage  probability limit, a full-diversity
capacity-achieving code is necessary, but may not be sufficient.

\section{Conclusions}\label{section:conclusions}
We have studied LDPC codes in the block-fading channel under both ML
and iterative decoding. We have shown that constructions designed for ML decoders
 fail to guarantee diversity under iterative decoding. Driven by this restriction, we have introduced the new family of root-LDPC codes,  which achieve full diversity under iterative decoding. We have shown both finite- and infinite-length performance, and we have illustrated how the error-rate performance of root-LDPC is close to the outage probability limit and almost insensitive to the block-length. This makes root-LDPC codes attractive for slowly-varying wireless communications scenarios.

\newpage
\appendices
\section{Coding gain of a $4$th-order
unbalanced $\chi^2$ distribution}
Here we limit our description to a diversity order of $2$, but all
results are easily extendable to rate-$1/n_c$ coding on a channel
with diversity order $n_c$. In the context of ML decoding, the
Euclidean distance between two codewords is proportional to
$\omega_1 \alpha_1^2 + \omega_2 \alpha_2^2$. As fading $\alpha_i$
have a Rayleigh density, their squares are exponentially
distributed, i.e., $p_{\alpha_i^2}(x)=e^{-x}$. The latter is a
central $\chi^2$ distribution of order $2$ with parameter
$\sigma^2=1/2$~\cite{Proakis2000}. Diversity $2$ is achieved with
a $\chi^2$ distribution of order $4$. Hence, a full-diversity code
must satisfy $\omega_1 >0$ and $\omega_1>0$ in order to get the
order-$4$, $\chi^2$ distributed, metric $\omega_1 \alpha_1^2 +
\omega_2 \alpha_2^2$. Once maximum diversity is guaranteed, the
maximization of the product $\omega_1 \omega_2$ increases the
coding gain.

\begin{figure}[htb]
\begin{center}
\includegraphics[width=0.99\columnwidth]{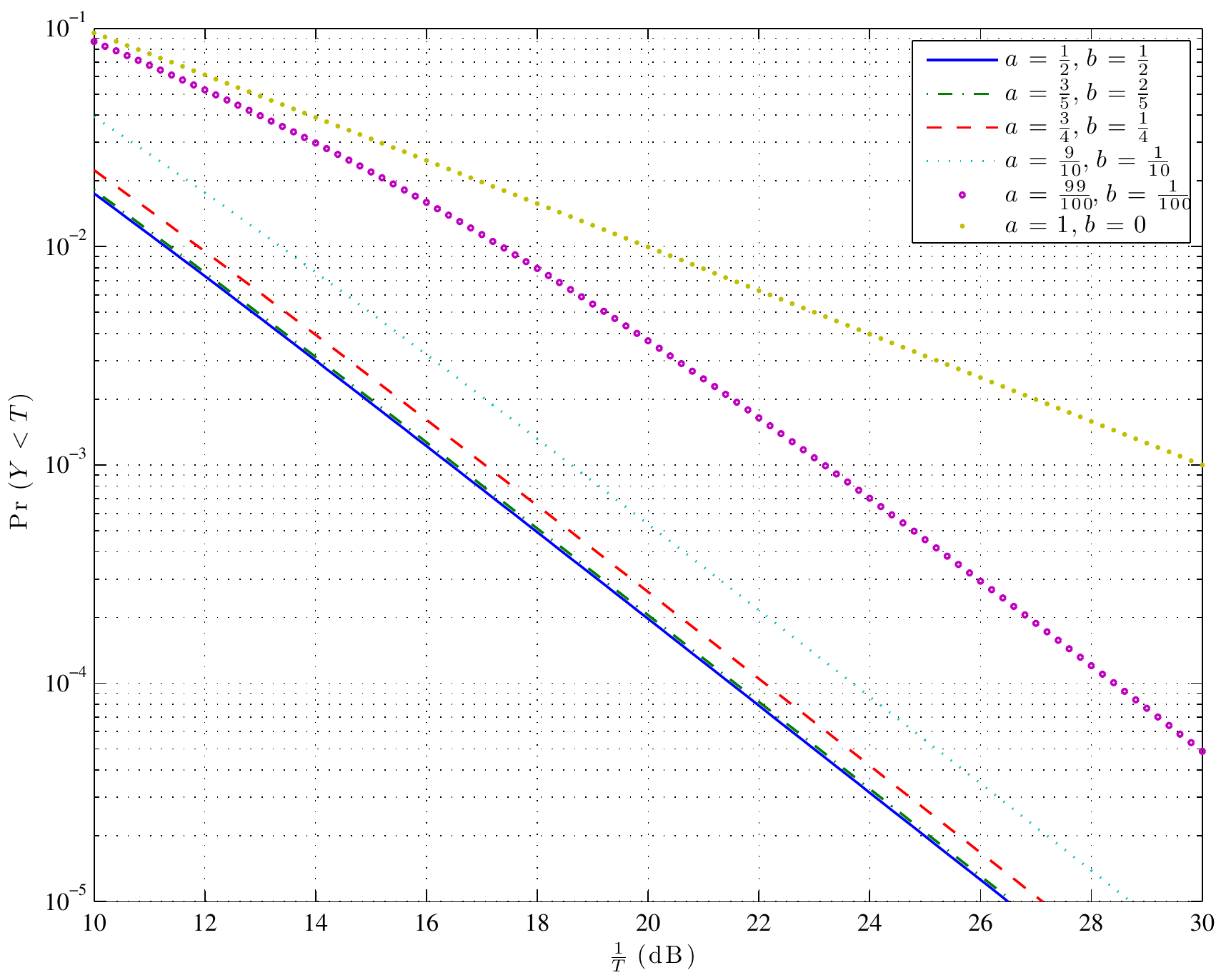}
\end{center}
\caption{\sl Coding gain and diversity order of $Y=a \alpha_1^2 +
b \alpha_2^2$ ($\chi^2$ of $4$th order) where $\alpha_1$ and
$\alpha_2$ are Rayleigh distributed.} \label{fig_coding_gain}
\end{figure}

The above simple facts are still valid in the context of iterative
probabilistic decoding. Let $\Lambda$ be the {\em a posteriori}
probability log-ratio of a binary element $b$. Achieving full
diversity under iterative decoding is equivalent to letting
$\Lambda$ behave as the metric $Y=a \alpha_1^2 + b \alpha_2^2$,
where $a$ and $b$ are two positive real numbers. The energy of $Y$
is normalized, $a+b=1$. The exact mathematical expression relating
$\Lambda$ to $Y$ depends on the type of iterative algorithm used
for decoding, e.g., $\Lambda \propto Y+\nu$ where $\nu$ is an
additive noise. To understand the influence of the product $ab$ on
the performance, one should study the error probability associated
with $Y$, i.e. $P(Y<T)=F(a,b,T)$. When $a=b=1/2$, the order-$4$
$\chi^2$ distribution is balanced, and its probability density
function is
\begin{equation}
p_Y(y) ~=~ 4ye^{-2y}
\end{equation}
When $a \ne b=1-a$, the order-4 $\chi^2$ distribution is
unbalanced, and its probability density function is
\begin{equation}
p_Y(y) ~=~ \frac{(e^{-y/a}-e^{-y/b})}{2a-1}
\end{equation}
The expression of $P(Y<T)=F(a,b,T)$ is obtained after integrating
$p_Y(y)$. The diversity order and the coding gain embedded in $Y$
appear when $T \ll 1$. For a balanced $\chi^2$ distribution, we
have
\begin{equation}
F(a,b,T) ~=~ 1-e^{-2T}(1+2T) ~=~ 2T^2 ~+~{\rm o}(T^2)
\end{equation}
For an unbalanced $\chi^2$ distribution, we obtain
\begin{equation}
F(a,b,T) ~=~ 1-\frac{ae^{-T/a}-be^{-T/b}}{2a-1} ~=~
\frac{T^2}{2ab} ~+~{\rm o}(T^2)
\end{equation}
In Fig. \ref{fig_coding_gain}, the performance function $F(a,b,T)$
is plotted versus $\gamma=1/T$ on a double logarithmic scale for
different values of $a$ and $b$. The slope is always $2$ (i.e.,
$F(a,b,T)\propto 1/\gamma^2$) for all positive values of $a$ and
$b$. The function $F$ degenerates to $T+{\rm o}(T)$ when $b=0$
(diversity order equal to $1$ instead of $2$). Notice also that an
unbalanced $\chi^2$ distribution with $a=3/4$ and $b=1/4$
generates a coding loss about $0.65$~dB. This loss is slightly
higher (about $0.75$~dB) when considering $P(\Lambda < 0)$ for
$\Lambda \propto Y+\nu$ since additive noise depends on the fading
coefficients as shown in Section~\ref{sectionIV}.
\begin{figure}[htb]
\begin{center}
\includegraphics[width=0.99\columnwidth]{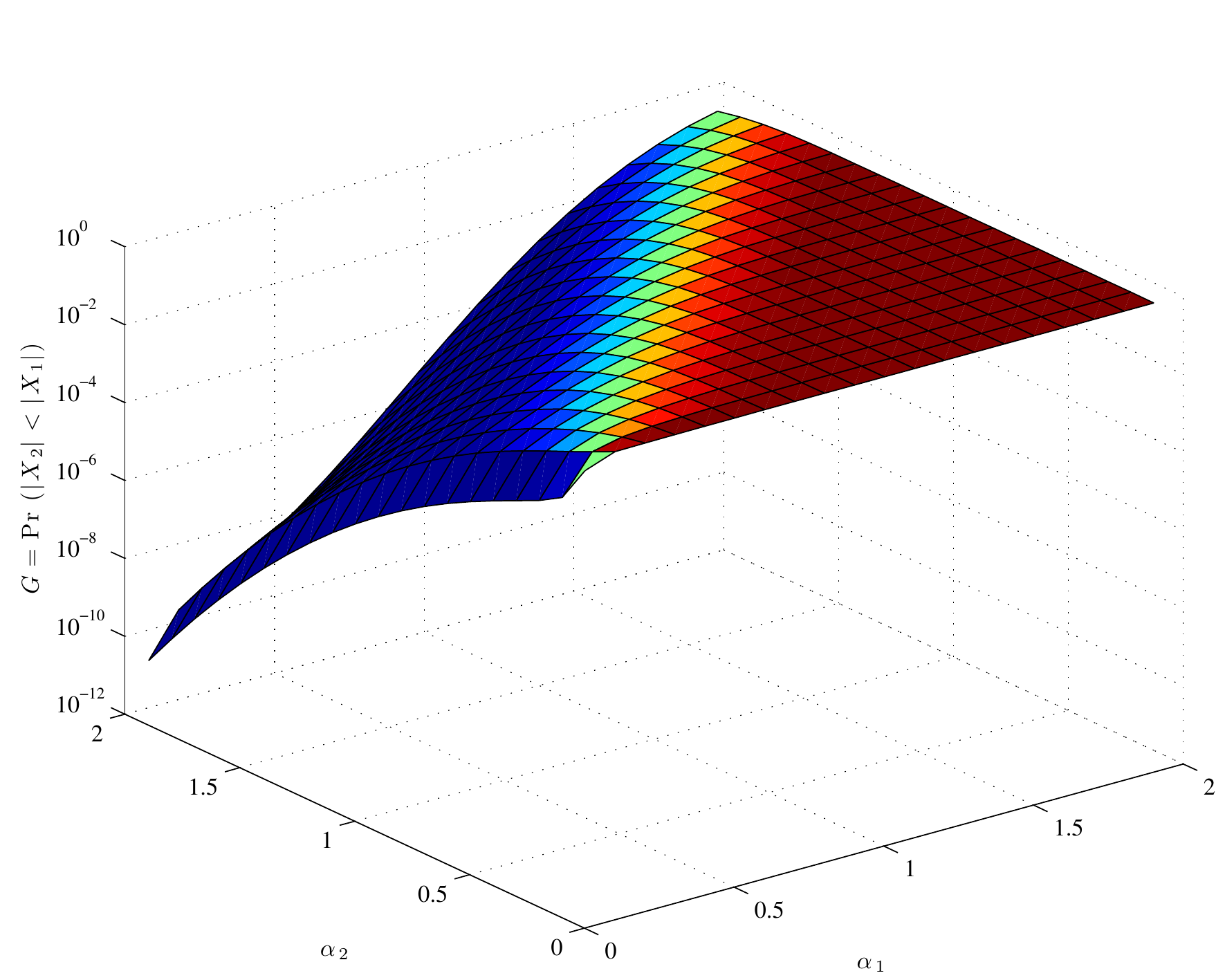}
\end{center}
\caption{\sl A 3D plot of $G={\rm Pr}(|X_2| < |X_1|)$ versus $\alpha_1$
and $\alpha_2$ for a variance $\sigma^2=1/10$.}
\label{fig_bidim_cdf2}
\end{figure}


\section{The bidimensional cumulative density function $G={\rm Pr}(|X_2| < |X_1|)$}
Consider two real independent Gaussian random variables $X_1 \sim
\mathcal{N}(\alpha_1^2, \alpha_1^2\sigma^2)$ and $X_2 \sim
\mathcal{N}(\alpha_2^2, \alpha_2^2\sigma^2)$. We define the
multivariate function $G(\alpha_1, \alpha_2, \sigma^2)\triangleq
{\mathbb P}(|X_2| < |X_1|)$. The $G$ function is given by the
integral expression
 \begin{equation}
G=1-\int_{0}^{\infty} \frac{dt}{\sqrt{2\pi \alpha_1^2\sigma^2}}
\left(e^{-\frac{(t-\alpha_1^2)^2}{2\alpha_1^2\sigma^2}} +
e^{-\frac{(t+\alpha_1^2)^2}{2\alpha_1^2\sigma^2}} \right) \left(
Q\left( \frac{t-\alpha_2}{\alpha_2\sigma} \right) + Q\left(
\frac{t+\alpha_2}{\alpha_2\sigma} \right)\right) \label{equ_G}
 \end{equation}
where $Q(x)$ is the Gaussian tail function. A 3D plot of $G$ is
illustrated in Fig. \ref{fig_bidim_cdf2}. The main properties of
$G$ are:
\begin{itemize}
\item $G(\alpha, \alpha, \sigma^2)=1/2$ for all $\sigma^2 > 0$.
\item $G$ is a non-decreasing function of $\alpha_1$ and a decreasing function of $\alpha_2$.
Hence, $G \le 1/2$ if $\alpha_1 \le \alpha_2$ and $G \ge 1/2$ if $\alpha_2 \le \alpha_1$.
\item For fixed $\sigma^2$ and $\alpha_2$, $G \rightarrow 1$ as $\alpha_1 \rightarrow +\infty$.
\item For fixed $\sigma^2$ and $\alpha_1$, $G \rightarrow 0$ as $\alpha_2 \rightarrow +\infty$.
\end{itemize}
%

\newpage
\begin{spacing}{1.3}

\end{spacing}

\end{document}